\documentclass[twocolumn,times,tighten]{aastex63}
\usepackage{amsmath}
\usepackage{graphicx}
\usepackage{xcolor}
\usepackage{float}
\usepackage{hyperref}

\usepackage{mathtools}
\usepackage[caption=false]{subfig}

\usepackage{hyperref}

\newcommand{\beq}{\begin{equation}}
\newcommand{\eeq}{\end{equation}}
\newcommand{\bea}{\begin{eqnarray}}
\newcommand{\eea}{\end{eqnarray}}

\newcommand{\rev}[1]{{\color{black}#1}}
\newcommand{\revtwo}[1]{{\color{black}#1}}
\newcommand{\revthree}[1]{{\color{black}#1}}

\begin{document}

\title{Using LSST Microlensing to Constrain Dark Compact Objects in Spherical and Disk Configurations}

\author{Harrison Winch}

\affiliation{David A. Dunlap Department of Astronomy \& Astrophysics, University of Toronto, 50 St. George St., Toronto, ON M5S 3H4, Canada}
\correspondingauthor{Harrison Winch}
\email{winch@astro.utoronto.ca}

\author{Jack Setford}
\affiliation{Department of Physics, University of Toronto, 60 St. George St., Toronto, ON M5S 3H4, Canada}

\author{Jo Bovy}
 \affiliation{David A. Dunlap Department of Astronomy \& Astrophysics, University of Toronto, 50 St. George St., Toronto, ON M5S 3H4, Canada}

\author{David Curtin}

\affiliation{Department of Physics, University of Toronto, 60 St. George St., Toronto, ON M5S 3H4, Canada}


\begin{abstract}
The Legacy Survey of Space and Time (LSST) with the Vera Rubin Observatory will provide strong microlensing constraints on dark compact objects (DCOs) in our Galaxy. However, most current forecasts limit their analysis to Primordial Black Holes (PBH). It is unclear how well LSST microlensing will be able to constrain alternative models of DCOs with different Galactic spatial profile distributions at a subdominant DM fraction. In this work, we investigate how well LSST microlensing will constrain spherical or disk-like Galactic spatial distributions of DCOs, taking into account extended observing times, baryonic microlensing background, and sky distribution of LSST sources. These extensions represent significant improvements over existing microlensing forecasts in terms of both accuracy and versatility. We demonstrate this power by deriving new LSST sensitivity projections for DCOs in spherical and disk-like distributions. We forecast that LSST will be able to constrain one solar mass PBHs to have a DM fraction under \revtwo{$4.1\times10^{-4}$}. One-solar-mass objects in a dark disk distribution with the same dimensions as the Galactic disk will be constrained below \revtwo{$3.1\times10^{-4}$}, while those with $m = 10^5M_{\odot}$ will be constrained to below \revtwo{$3.4\times10^{-5}$}. We find that compressed dark disks can be constrained up to a factor of $\sim10$ better than ones with identical dimensions to the baryonic disk. We also find that dark disks become less tightly constrained when they are tilted with respect to our own disk.  This forecasting software is a versatile tool, capable of constraining any model of DCOs in the Milky Way with microlensing,   \href{https://github.com/HarrisonWinch96/DarkDisk_Microlensing}{\textcolor{blue}{and is made publically available}}.
\end{abstract}

\section{Introduction}\label{Sec:Intro} 

Microlensing is a phenomenon whereby a massive object passes directly in front of a bright background source, gravitationally lensing the light \citep{ML_Original_Paczynski:1985jf, MACHO_Original_results_Alcock:1995zx}. However, the Einstein ring created by the strong lensing in these cases is too small to be resolved by most telescopes, so the event manifests only as a temporary brightening of the background object. This temporary brightening can be detected with long-term photometric survey telescopes, and can be distinguished from other transients by its achromatic and time-symmetric lightcurve. Characterizing the number of microlensing events we see in a given survey can put constraints on the number of dark compact objects in our Galaxy, allowing us to constrain models of compact DM. This technique has been used in the past to rule out MAssive Compact Halo ObjectS (MACHOS) for a variety of mass ranges comprising 100\% of the DM in our halo \citep{MACHO_Original_results_Alcock:1995zx, OGLE_ML_Wyrzykowski:2015ppa}. Future surveys, such as the Legacy Survey of Space and Time (LSST) at the Vera C. Rubin Observatory, should be able to push the constraint on MACHO DM fraction much lower due to its increased sensitivity and number of target sources \citep{LSST_MLPredictions, LSSTDM}. It will also be able to use paralensing (\rev{a periodic feature imposed on the lensing signal due to parallax effects}) to detect microlensing events with arbitrarily long crossing times \citep{Gould92a}. \rev{Paralensing has been observed in past surveys \citep{Wyrzykowski_2016, golovich2020reanalysis}, but never with the large number of sources available to LSST. }

\begin{figure*}
    \centering
    \includegraphics[width=0.8\textwidth]{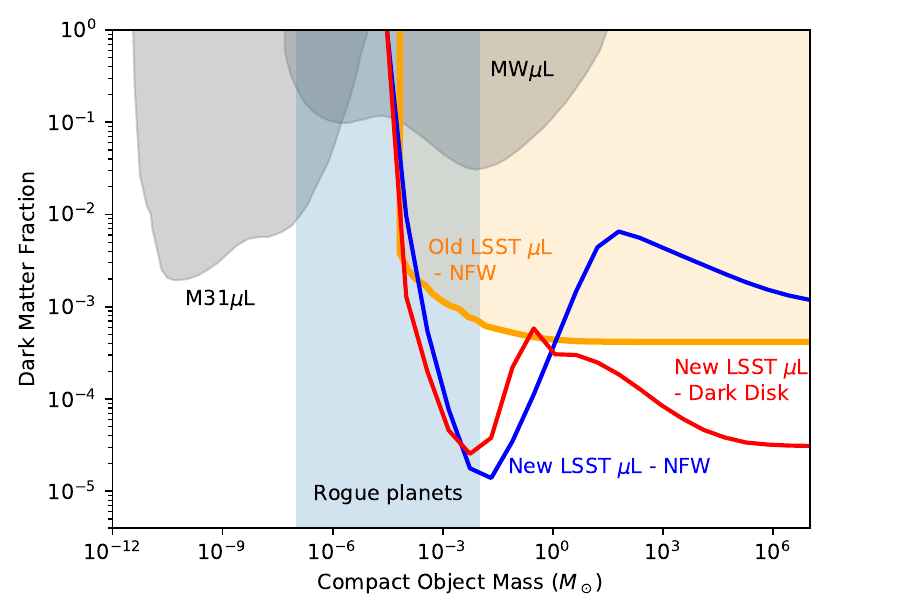}
   \caption{Past and future constraints on the DM fraction of dark compact objects in our Galaxy as a function of object mass, with shaded regions being constrained.  The MW$\mu$L curves are from the MACHO and OGLE surveys of the LMC and Galactic bulge \citep{Alcock_2000, Tisserand, Wyrzykowski_2009}. The M31$\mu$L curve is from detecting microlensing in stars in the M31 galaxy \citep{Niikura:2017zjd}. The orange curve represents current forecast microlensing constraints from LSST, as described in \citet{LSSTDM}. The blue curve shows the new constraints calculated in this paper, using full observing runtime, baryonic microlensing background rates, multiple lines of sight, and variable lens velocity distributions. Both the orange and blue curves assume a spherically-symmetric NFW distribution of dark compact objects \citep{Navarro_1996}. \rev{See Section \ref{sec:results} for a more details on this computation.} The red curve shows the forecast constraints on dark compact objects in a benchmark dark disk distribution, parameterized as a fraction of total dark matter. This dark disk has a double exponential profile with the same size and shape as our Galactic disk, but with a delta function mass distribution. See Section \ref{sec:ad_mirror_disk} for details and constraints for generalized dark disks. The pale blue band indicates the mass range where unknown numbers of rogue planets may exist in interstellar space, creating an irreducible systematic that would impede searches for dark compact objects in that range. \rev{Note that the Roman space telescope will eventually impose constraints on this plot as well (as discussed in \citet{Johnson_2020}), but that these constraints will likely not surpass those from LSST.}}  
    \label{fig:old_constraints}
\end{figure*}

Fig. \ref{fig:old_constraints} shows forecast constraints on the compact object halo fraction as a function of object mass for spherically-distributed compact DM, and it is clear that LSST will do much better than past experiments detecting microlensing in either the Milky Way (MW$\mu$L) or the Andromeda Galaxy (M31$\mu$L) \rev{(based on past microlensing surveys \citet{Alcock_2000, Tisserand, Wyrzykowski_2009, Niikura:2017zjd})}. The past LSST forecasts (in orange) are taken from \citet{LSSTDM}. The new forecast for this compact DM distribution (in blue) are one subject of this paper.

It is not obvious how these constraints would translate onto alternative theories of compact dark matter; this is due to the fact that these forecasts, \cite{LSSTDM} assume that the compact objects being investigated by LSST are Primordial Black Holes (PBH). Primordial Black Holes have a strong theoretical justification (only requiring minor adjustments to early-universe physics, and minimal extensions to the standard model), and their existence has been suggested as an explanation of some of the higher-mass binary black hole pairs detected by LIGO, thus it is understandable for past microlensing analysis to focus on them as the primary compact DM candidate \rev{\citep{LIGO_IMBH, dolgov2019massive, Jedamzik:2020ypm}}\footnote{\rev{For a more thorough explanation of PBHs and their origins and theoretical justifications, we refer the reader to the work cited here.}}. PBHs are assumed to have a spherically symmetric and \rev{NFW} Galactic distribution, as they are practically collisionless. This assumption, along with several other simplifications made in the calculations of \citet{LSSTDM}, play a significant role in deriving the LSST sensitivity to the dark object DM fraction. Therefore, alternative models of compact dark matter -- beyond the leading PBH candidate -- would be constrained differently by the same survey. 

Particularly interesting are models of \emph{dissipative} dark matter, i.e., dark matter that emits dark radiation and can therefore cool efficiently. In most models dissipative dark matter is constrained to be only a subcomponent, making up at most around 5-10\% of the  total dark matter density \citep{Fan:2013yva,Cyr-Racine:2013fsa,Chacko:2018vss}. Dissipative dark matter can form radically different structures compared to ordinary cold, collisionless dark matter; for instance, it can cool to form a \emph{dark disk} in analogy to the baryonic disk of our own Galaxy \citep{Fan:2013yva, Fan:2013tia, Kramer:2016dew, Kramer:2016dqu}.

Within such models, \emph{compact} dark objects can form --- see \citet{Chang:2018bgx} for a simple model featuring only an asymmetric dark electron and a dark photon. The size and mass of these compact objects depends on the details of the model, being set essentially by the self-interactions and cooling rate of the dissipative subcomponent, as well as possible sources of feedback. In principle. any such model can be constrained by microlensing observations, and LSST will be able to offer the best constraints yet for a broad region of the parameter space.\footnote{Other gravitational probes of dark disks include measurements of the local matter density, which can constrain extremely thin dark disks with thickness $< 100~\mathrm{pc}$ to contain \rev{$<1\%$ of the Milky Way's DM}~\citep{Gaia_thindisk, Buch:2018qdr}}

One particularly interesting example of an alternative form of compact DM arises in the Mirror Twin Higgs framework \citep{MTH_Chacko:2005pe} with asymmetric reheating \citep{Chacko:2016hvu,Craig:2016lyx}. This model is representative of a broad class of ``neutral naturalness'' solutions to the little hierarchy problem, which refers to the fact that quantum corrections destabilize the mass of the Higgs boson in the Standard Model (SM) of particle physics, motivating the existence of new particles to cancel those quantum corrections (see ~\cite{Martin:1997ns,Csaki:2018muy} for a review).
As opposed to theories such as low-energy supersymmetry, theories of neutral naturalness do not feature new particles with TeV-scale masses that carry SM colour charge; as a result, they are largely unconstrained by LHC null results to date. 
The Mirror Twin Higgs hypothesis proposes a separate mirror sector of particles, similar to our own SM. Due to a larger Higgs Vacuum Expectation Value (VEV) in the mirror sector, fundamental particles in the mirror sector would all be more massive than their SM counterparts, by a factor of between $\sim$3 and 7 in the simplest and most natural models. 
This mass enhancement is a free parameter of the model, that can be constrained either by Higgs decay measurements in collider experiments \citep{Burdman:2014zta} or by searching for astrophysical signatures of mirror baryons \citep{MTH_Cosmology_Chacko:2018vss}. These mirror sector particles mostly interact with themselves, being endowed with the dark equivalents of electromagnetism and nuclear physics, and acting effectively like a small self-interacting sub-component of DM.

Some work has been done calculating the cosmological implications of such a mirror sector, particularly in \cite{MTH_Cosmology_Chacko:2018vss}, in addition to earlier work on fully symmetric mirror DM such as \cite{DM_historical1}. However, we are only beginning to understand the astrophysical implications of this model. Depending on the model parameters, these baryon-like DM particles can form dark galactic disks and coalesce into ``mirror stars'' that fuse mirror nuclei and shine in mirror light. These mirror stars would be cohabiting our halo but be largely invisible to us, except for the possibility of faint characteristic optical and X-ray signatures in the presence of mixing between the mirror and SM photons \citep{Curtin:2019ngc, Curtin:2019lhm}. (For earlier work on the possibility of dark matter stars, see \citep{Mohapatra:1996yy, Mohapatra:1999ih, Foot:2000vy, Berezhiani:2003xm, Berezhiani:2005vv}.) These mirror stars would act like a form of compact dark matter, albeit one with significantly different properties than PBHs.
LSST should be able to probe the existence and properties of such a mirror star population in our Galaxy, independent of any electromagnetic signals they may or may not generate.
However, the different mass and spatial Galactic distributions of mirror stars compared to PBH's  change how well LSST will be able to constrain the fraction of our DM halo consisting of mirror stars with microlensing. \rev{The details of how the properties of the mirror-sector model would connect to the mass and spatial distributions of the mirror stars are complicated and as-yet unsolved \citep{Curtin:2019lhm}. Therefore, it is important to formulate model-independent microlensing constraints on DCOs in Galactic distributions that are more general than that of PBHs, so that they might eventually be applied to a variety of Dark Disk models.}

Other exotic compact objects that can appear in new physics models are boson stars \citep{Tkachev:1991ka, Hogan:1988mp, Kolb:1993zz}, including axion stars \citep{Kolb:1993zz, Barranco:2010ib,Eby:2014fya}. A wide range of masses are possible for such objects, depending on the parameters of the model in question. We also direct the reader to a recent study on the effect of size and density distribution of extended structures on microlensing signals \citep{Croon:2020wpr, Croon:2020ouk}.

Given the variety of dark sector models that can give rise to exotic compact objects, it would be helpful to anticipate how well LSST can constrain dark compact objects with an arbitrary Galactic spatial and mass
distribution. This would give a flexible way to calculate constraints that could change depending on the uncertainties in our understanding of mirror sector astrophysics, or other models of dark compact objects. 
In this paper, we produce numerical forecasts for LSST microlensing that take as input an arbitrary functional form for the Galactic spatial distributions of dark compact objects, and predict the constraints LSST will be able to impose on the DM fraction. Our software is made publicly available at \cite{harrisongithub}, and can be used to compute forecast constraints for any arbitrary spatial distribution of dark compact objects. These constraints also include a number of systematic improvements over the initial forecasts by \citet{LSSTDM} by including variable survey length, background baryonic microlensing events, and the contribution from multiple lines of sight. Note that these constraints might be subject to change, depending on the final choice of observing strategies decided upon by the LSST collaboration; further discussion on the impact of observing strategies can be found in \citet{LSST_MLPredictions}.
In addition to being very helpful for the development of mirror-sector astrophysics (being able to anticipate the constraining power for different Galactic and stellar models), this code is generally applicable for any theory of complex self-interacting DM that could produce dark compact objects.

In the future we will extend this software to also provide constraints on arbitrary mass distributions,
similar to the work done in \citet{PBH_EinsteinTime_Lu:2019hoc}.
This could result in weaker constraints on the DM fraction, particularly if the DCO mass distribution is similar to the stellar mass distribution (which would result in sensitivities that are limited by our knowledge of stellar abundances). A more complete discussion of the potential impact of mass distributions is given in Sec. \ref{sec:fut_app}. \rev{Another possible extension to this work is to analyze the impact of DCO clustering on microlensing constraints, similar to the work done in \cite{Clesse_2017}.}
%
%

One potential difficulty in detecting dark objects in the $\sim 10^{-3} M_\odot$ range is that past microlensing surveys have suggested the presence of wandering Jovian planets in our Galaxy, which would be indistinguishable from dark compact objects \citep{planet_mL_1, planet_mL}. Until we better understand this population of Jovian lenses, the prospects for positively detecting dark objects in this mass range are uncertain. \rev{We should also note that the mass range most impacted by rogue Jovian planets (from roughly $10^{-7}$ to $10^{-2} M_\odot$) is also the mass range most impacted by changes in the observing cadence or sensitivity of LSST, adding to the uncertainties in this range.}

In Section \ref{sec:methods}, we explain how the microlensing constraints were calculated from theoretical microlensing rates. We also explain several of the novel improvements to our calculations of microlensing sensitivity, such as the inclusion of extended observing times in \ref{sec:rates_and_times}, baryonic microlensing background events in \ref{sec:baryons} and \ref{sec:probabilisticcomparison}, realistic velocity distributions in \ref{sec:velocities}, and combined constraints from multiple lines of sight in \ref{sec:lsst_los}. In Section \ref{sec:results}, we show the effect that including each of these improvements has on our forecasted microlensing constraints for spherical distributions of dark compact objects. Finally, in Section~\ref{sec:alt_dens}, we discuss constraints on alternative distributions of dark compact objects, such as a flattened halo in \ref{sec:ad_flat_halo}, a dark disk rescaled with respect to our own baryonic disk in \ref{sec:ad_mirror_disk} that may apply to mirror stars, and a dark disk that is tilted with respect to our own in \ref{sec:ad_tilt}. Section \ref{sec:fut_app} presents some possible trivial and non-trivial extensions to this work using the software. Final conclusions are presented in Section~\ref{sec:conclusion}.

\section{Methods: Microlensing Computations}
\label{sec:methods}

In order to put constraints on the quantity of dark objects in the Galaxy, we need to estimate the average number of microlensing events by dark compact objects that should be seen, given an assumed density of dark compact objects, and then calculate the probability of observing some actual number of observed events. This actual number could be zero, in the case of no positive events, or simply the estimated number of baryonic microlensing events caused by baryonic compact objects (also known as stars and planets). We can then determine the maximum dark matter fraction that could be consistent with some fiducial result in order to estimate the resulting constraints.
We set the mass distribution of dark compact objects to be delta functions in the calculations that follow, both for simplicity and for comparison to earlier calculations. This overestimates the sensitivity of microlensing to small dark matter fractions compared to more \rev{realistically spread-out} mass distributions, but we leave a quantitative analysis of LSST sensitivity to \rev{continuous} mass distributions of DCOs for future study.

\subsection{The expected number of microlensing events in a survey}
\label{sec:rates_and_times}

In \cite{LSSTDM}, the expected number of microlensing events was calculated using only the optical depth to microlensing, which estimates the number of sources undergoing a microlensing event at any single point in time. However, this calculation does not take into account the full survey time, as short events could experience more detections if a survey lasts for $\sim 12$ years, as LSST is expected to. In order to calculate the expected number of microlensing events over an extended period of time ($N_\mathrm{expected}$), we need to calculate the differential microlensing event rate per unit crossing time, multiply by the detection efficiency for events with that crossing time ($\xi(\hat{t})$), multiply by the range of times over which an event of a given crossing time could start and we would still detect it, and then integrate over crossing time. This gives some total number of microlensing events observed per source, which can then be scaled to the \revtwo{$\approx 4$ billion} stars anticipated to be seen by LSST \revtwo{in each observation \citep{LSSTDM, LSSTScieneBook}} \footnote{\rev{Technically, microlensing can occur with either a star or a quasar as the source. However, for our analysis, we have used the expected number of stars resolvable by LSST (4 billion), and neglected the number of quasars. This is because LSST is expected to observe only 10 million quasars, representing a correction of less than one permille to our constraints \citep{ivezic_2016, LSSTScieneBook}}}. This technique can be used to estimate both the number of dark and baryonic microlensing events, using different mass and Galactic distribution functions.

The differential microlensing event rate, which describes the number of microlensing event rates per unit time per unit crossing time ($\hat{t}$, how long the microlensing event lasts), is given in \cite{MACHO_Original_results_Alcock:1995zx} and \cite{MACHO_Update_mass_formalism_Calcino:2018mwh} to be
\begin{equation}
\label{dGammadt}
\frac{d\Gamma}{d\hat{t}} = \frac{32 L}{\hat{t}^4 m v_c^2}\int_0^1 \rho_{\mathrm{DM}}(x) r_E^4(x,m) \exp\bigg(-\frac{4 r_E^2(x,m)}{\hat{t}^2 v_c^2}\bigg) dx,
\end{equation}
where $\hat{t}$ is the event crossing time, $L$ is the distance from the observer to the source (for stars in the bulge, we use $L = 8.2$ kpc \citep{GRAVITY}), $m$ is the object mass, $v_c$ is the relative velocity of the DM halo with respect to us (which, for an isotropic DM halo, we set to $v_c = 220\,\mathrm{km\,s}^{-1}$ \citep{vc_Bovy_2012}), $x$ is the fractional distance of the lens compared to the source, $\rho_\mathrm{DM}(x)$ is the compact DM density at that distance along the line of site (which would be linearly resclaed by $f_{DM}$ in the case of a subdominant compact DM fraction), and $r_E(x,m)$ is the Einstein radius for a given distance and mass. The derivation of this formula can be found in \cite{MACHO_Original_results_Alcock:1995zx} and \cite{griest_originalderivation}. The halo density profile used in the derivation of the PBH constraints is the Navarro-Frenk-White (NFW) profile \citep{Navarro_1996} of the form
\begin{equation}\label{eq-nfw}
    \rho_\mathrm{NFW}(r) = \frac{\rho_0}{\big(\frac{r}{r_s}\big)\big( 1 + \frac{r}{r_s}\big)^2}
\end{equation}
with the parameters of $\rho_0 = 0.014$ M$_\odot$ pc$^{-3}$ and $R_s = 16$ kpc \citep{MW_halo_dist,localDM_Bovy:2012tw,starvel_Bovy_2015}. The Einstein radius is given by the standard formula
\begin{equation}
r_E = \sqrt{\frac{4 m L G x (1-x)}{c^2}}.
\end{equation}

 \rev{This microlensing rate must then be multiplied by the range of possible start times for an event to be observable. This range of start times depends on whether the event is being detected using traditional microlensing, or paralensing, which is the oscillatory lensing amplification caused by the Earth's orbit instead of the transverse velocity of the object itself. For traditional microlensing, the event must fully overlap with the survey in order to be considered detectable, as we would need to see both the rising and falling sections of the light curve to confirm that the event is achromatic and time symmetric. Therefore, the range of possible event start times that could be detectable would be $T_\mathrm{micro} = t_\mathrm{survey} - \hat{t}$. However, for paralensing events,} the event can start or end before or after the actual observing period, as long as it overlaps with the observing period \rev{for at least one year, in order to detect one full parallax oscillation}. Therefore, the length of time over which an event of duration $\hat{t}$ can start and still be seen \rev{by paralensing is $T_\mathrm{para} = t_\mathrm{survey} + \hat{t} - (2 \; \mathrm{ years})$. This includes both events that start during the survey (but not in the final year of the survey), and events that start up to $\hat{t} - (1 \; \mathrm{year})$ before the survey (so they end at least one year after the survey begins, providing the necessary overlap). Therefore, for a given event duration $\hat{t}$, the range of possible start times for it to be detectable by either microlensing or paralensing is the greater of the two ranges provided above. For short-duration events, this would be traditional microlensing, while longer events would be dectable with paralensing. }The total number of expected microlensing events is then,
\begin{equation}
    N_\mathrm{observed} = n_\mathrm{sources} \int_0^\infty \frac{d\Gamma}{d\hat{t}} T_\mathrm{micro/para}\xi(\hat{t}) d\hat{t}
\end{equation}

Using this method for calculating microlensing event numbers, we can compute the constraints on compact object DM fraction.

\rev{The detectability of events with crossing time $\hat{t}$ was modeled by \cite{LSSTDM} using well-informed assumptions about cadence and sensitivity of the LSST experiment. Since the sensitivity of the LSST instrument and analysis pipeline is outside the scope of this work, we used an analytic function to describe this detectability, $\xi(\hat{t})$, with parameters fit to match the mass-dependant sensitivity presented in \cite{LSSTDM}. We used a logarithmic logistic function of the form
\begin{equation}
    \xi(\hat{t}) = \frac{1}{1 + \big( \frac{\hat{t}}{t_0} \big)^{-1/t_r}}.
\end{equation}
This function was chosen because it varies smoothly with $\log \hat t$ between complete detectability and no detectability, and gives a reasonable fit to the results of~\cite{LSSTDM}. 
As an additional check, we confirm that the we can reproduce the forecast from  \cite{LSSTDM}, shown as the orange curve in Figure \ref{fig:old_constraints}, using our detectability function and their optical-depth-only approach. This gives us confidence that this detectability function is sufficient  for our more realistic sensitivity forecasts as well.


}
\subsection{Baryonic microlensing comparison}
\label{sec:baryons}

The calculations in \cite{LSSTDM} assume that we will observe zero microlensing events with LSST, and then calculate the DM fraction required to make the predicted number of events fall below one. However, in reality, we should observe many microlensing events, simply from ordinary baryonic stars crossing the lines of sight to background sources. Taken in isolation, these events would be indistinguishable from microlensing events caused by compact dark objects. This is due to the fact that the lens star would be so close to the background source star that they would likely be unresolvable with the telescope used (otherwise the lensing event would be classified as strong lensing and not microlensing). \rev{Although it would be theoretically possible to conduct a spectroscopic or multi-band analysis to determine the presence of two stars, we make the conservative assumption that this analysis is not done. Barring such analysis, t}he only way to distinguish the dark and baryonic microlensing events is by computing their statistical distribution (in both crossing time $\hat{t}$ and location on the sky $\vec{y}$) and comparing it to the distribution expected from baryonic stars. One microlensing event is not enough to be a detection --- we can only claim to have detected a number of dark object microlensing events if that number is greater than the uncertainty on the number of baryonic microlensing events of the same crossing time. This uncertainty must incorporate both inherent Poisson uncertainty on the number of baryonic microlensing events, and measurement uncertainty on the number of stars in our Galaxy, which we estimate to be roughly a 5\% uncertainty on the number of stars in any given mass bin, consistent with \cite{stardense_Bovy_2017}. 

In order to calculate the baryonic microlensing event rate, we use the same procedure for calculating the dark object event rate discussed in Section \ref{sec:rates_and_times}, replacing the dark object density profile with a simple double-exponential Galactic disk model
\begin{equation}
\label{rhostar}
    \rho_\mathrm{star}(R,z) = A_b \exp\bigg(-\frac{(R - R_0)}{h_R}\bigg) \exp\bigg(-\frac{|z|}{h_z}\bigg)
\end{equation}
where $A_b = 0.04 M_\odot\,\mathrm{pc}^{-3}$ is the local stellar density \citep{stardense_Bovy_2017}, $R_0 = 8.2$ kpc is the Sun's distance from the Galactic center, and $h_R = 3$ kpc and $h_z = 400$ pc are the radial and vertical scale lengths, respectively \citep{starvel_Bovy_2015}. We also must take into account the mass distribution of baryonic stars, using a normalized Kroupa initial mass function (IMF) \citep{Kroupa_IMF} of 
\begin{equation}
    N_*(m) = C m^{-\alpha},
\end{equation}
where
\begin{equation}
\alpha = 
    \begin{cases}
    0.3\,, & m \leq 0.08\,M_\odot \\
    1.3\,, & 0.08\,M_\odot < m \leq 0.5\,M_\odot \\
    2.3\,, & 0.5\,M_\odot < m
    \end{cases}.
\end{equation}
\rev{This distribution is intended to reflect the population of both stellar lenses and compact stellar remnants, which are still produced according to the stellar IMF and could act as non-luminous lenses. 
It is of course the case that compact stellar remnants will be less massive than the high-mass stars that created them, meaning the IMF overestimates the abundance of high-mass objects compared to the real present-day mass distribution. (On the other hand, the stellar remnant contribution to lower mass populations is neglibible in relative terms.) 
%
This overestimation of the baryonic lens population means that our projections for the LSST sensitivity to dark compact objects must be regarded as a conservative estimate, and the real constraints may be slightly more stringent in the high-stellar-mass range. 


The distinction between luminous and non-luminous baryonic lenses is not relevant for our work, as the resulting microlensing events would only be distinguishable via spectroscopic or multi-band analysis, which we assume is not done, though it could be the subject of dedicated additional observations. If such an analysis were conducted, some of the microlensing events from luminous baryonic lenses could be positively identified, potentially tightening the resulting constraints on DCOs. However, the feasibility of such an analysis is outside the scope of this work, so we defer to the conservative assumption that no attempt to spectrally identify baryonic lensing events is conducted. Therefore, microlensing events from DCOs, compact stellar remnants, and luminous stars are all considered indistinguishable, with the only defining features being the event crossing time and location on the sky.}



\subsection{Probabilistic comparison of Dark and Baryonic Microlensing}
\label{sec:probabilisticcomparison}

Next we need to calculate the probability of a certain dark matter fraction given a certain observed number of events, and the theoretical expectation of how many dark and baryonic events we should see. We also need to incorporate our uncertainty on the total amount of baryonic stars in the Galaxy. Let us first assume that our basic model of the observations is characterized by two parameters: $f_\mathrm{DM}$, that sets the fraction of DM in compact objects in our assumed  distribution, and a parameter $\alpha_b$, that scales the expected baryon density, with $\alpha_b = 1$ corresponding to our fiducial baryon model from equation~\eqref{rhostar} above. We also assume a prior constraint on $\alpha_b$ that it has an uncertainty of 5\%, that is, $p(\alpha_b) = \mathcal{N}(\alpha_b|1,0.05^2)$, where $\mathcal{N}(x|\mu,\sigma^2)$ is the Gaussian distribution for $x$ with mean $\mu$ and variance $\sigma^2$. \rev{We assume a uniform prior on $f_\mathrm{DM}$ between 0 and 1, so $p(f_\mathrm{DM}) = 1$ on $0 \leq f_\mathrm{DM} \leq 1$, and $=0$ elsewhere}. We use a simulated set of microlensing events calculated using the method above, consisting of event rates for $N_{LoS}$ lines of sight, each representing $n_{s}$ number of actual sources. For a given simulated set of data consisting of numbers of microlensing events $D = \{N(\hat{t}^*, \vec{y})\}$ observed in bins in duration $\hat{t}^*$ and line-of-sight $\vec{y}$, the posterior distribution is
\begin{equation}
\begin{split}
   p( f_\mathrm{DM},\alpha_b) |& \{N(\hat{t}^*, \vec{y})\} ) \propto  \\ &p(\{N(\hat{t}^*, \vec{y})\} | f_\mathrm{DM},\alpha_b) \,p(\alpha_b) \,p(f_\mathrm{DM})\,.
\end{split}
\end{equation}
where the likelihood is given by
\begin{equation}
\begin{split}
        p(& \{N(\hat{t}^*,  \vec{y})\} | f_\mathrm{DM},\alpha_b) = \\ &  \prod_{\hat{t}^*, \vec{y}} \big[ \mathrm{Poi}(N(\hat{t}^*, \vec{y})|\left[N_{\mathrm{DM}}(\hat{t}^*, \vec{y};f_\mathrm{DM})+\alpha_b\,N_{\mathrm{star}}(\hat{t}^*, \vec{y})\right]) \big]^{n_{s}}\,,
\end{split}
\end{equation}
$N_{\mathrm{DM}}(\hat{t}^*, \vec{y};f_\mathrm{DM})$ is the number of dark matter events predicted for the given $f_\mathrm{DM}$, and $N_{\mathrm{star}}(\hat{t}^*, \vec{y})$ is the number of baryonic events for the fiducial model (and is therefore scaled by $\alpha_b$). The power of $n_{s}$ represents the fact that this probability is being computed for $n_{s}$ individual sources along each line of sight, but that we are approximating the event rates as all being equal to $N(\hat{t}^*, \vec{y})$. We thus take the product of all the likelihoods, corresponding to the joint likelihood from all the real sources that each line of sight represents. Marginalizing the posterior distribution over $\alpha_b$ then gives 
\begin{equation}
\begin{split}
    p( f_\mathrm{DM}| \{N(\hat{t}^*, \vec{y})\} ) 
    \propto p(f_\mathrm{DM})&\,\\ \times \int\mathrm{d}\alpha_b\,p(\alpha_b)\prod_{\hat{t}^*, \vec{y}} \big[\mathrm{Poi}(N&(\hat{t}^*, \vec{y})| [N_{\mathrm{DM}}(\hat{t}^*, \vec{y};f_\mathrm{DM}) \\ &+\alpha_b\,N_{\mathrm{star}}(\hat{t}^*, \vec{y}) ]) \big]^{n_{s}}.
\end{split}
\end{equation}

Constraining $f_\mathrm{DM}$ to a given confidence level $C$ (e.g., $C=0.95$ for 95\% confidence $\approx 2\sigma$) then requires us to (a) simulate a set of events $\{N(\hat{t}^*, \vec{y})\}$ assuming that $f_\mathrm{DM} = 0$ and $\alpha_b=1$ and (b) for a given DM distribution, find $f_\mathrm{DM}$ such that $\int \mathrm{d}f_\mathrm{DM}\,p( f_\mathrm{DM}| \{N(\hat{t}^*, \vec{y})\} ) =C$.

\subsection{Dark Compact Object Velocity Distributions}
\label{sec:velocities}

In addition to having a different density distribution function from the DM, microlensing with baryonic stars or a disk of dark compact objects also involves a different velocity distribution function because we can no longer assume that the lens velocities are isotropic, but are instead co-rotating along with us. This requires more complicated calculations of lens velocities.

For this first treatment of dark compact objects in disk-like distributions, we assume that the circular velocity and velocity dispersion profiles are identical to those of the Milky Way. This should be a reasonable approximation for dark disks that are subdominant in mass to the baryonic disk. The rotation curves are generated using the Galactic dynamics Python code \texttt{galpy} \citep{starvel_Bovy_2015}. Thus the average velocity at a given radius from the Galactic center is fixed by the object's position (and the angular tilt of the dark disk). In this case the relative velocity $v_c$ to be used in equation \eqref{dGammadt} is the relative velocity of the lens and source transverse to the line of sight, that is,
\begin{equation}
    v_c = |\vec v_{source}^\perp - \vec v_{lens}^\perp|,
\end{equation}
where the superscript $\perp$ indicates the component transverse to the line of sight connecting the Earth and the source. Since both the source and the lens can be co-rotating with the sun with a similar circular velocity, $v_c$ can be significantly less than the average value of $220\,\mathrm{km\,s}^{-1}$ that one would expect for an isotropic dark matter halo profile.

\subsection{Simulating LSST sources}
\label{sec:lsst_los}

\begin{figure*}
    \centering
    \includegraphics[width=\textwidth]{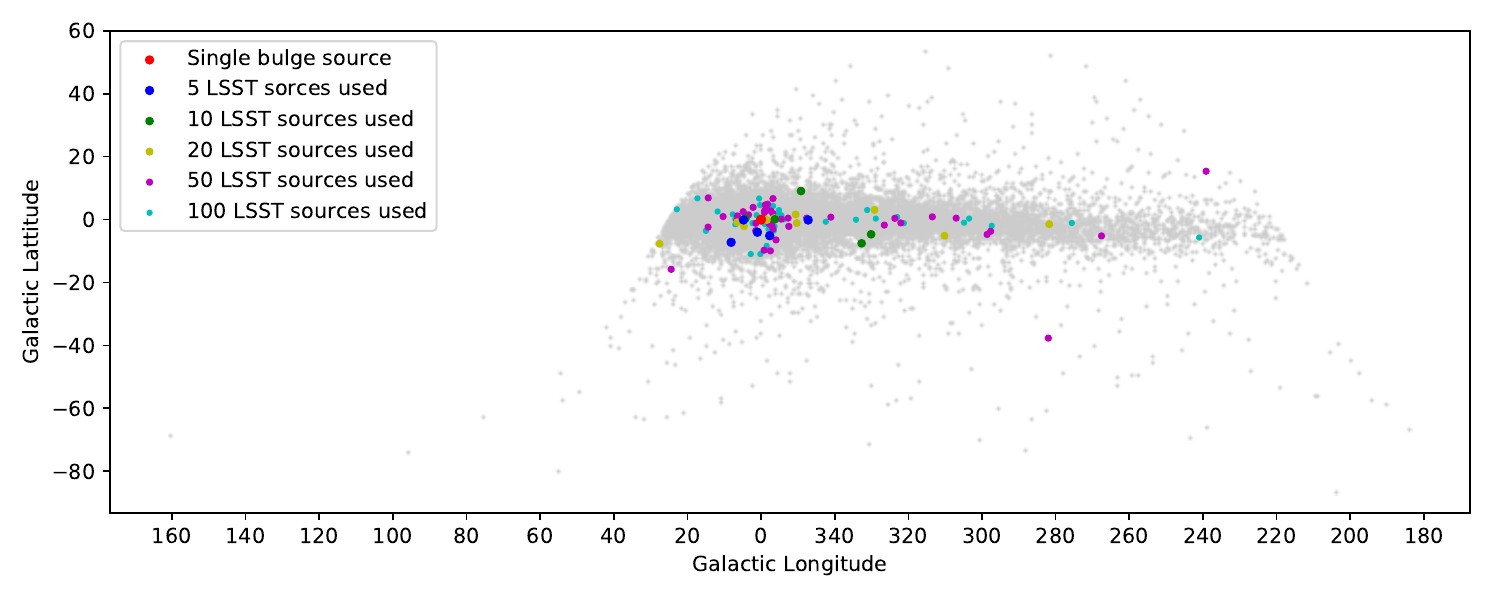}
   \caption{Sky distribution in Galactic coordinates of the simulated LSST \rev{stars} generated using Galaxia, along with the randomly chosen 5, 10, 20, 50, and 100 sources used to calculate microlensing event rates. It is clear that the locations of these random \rev{stars} is fairly representative of the total coordinate distribution of LSST objects.} 
    \label{fig:sourcedistribution}
\end{figure*}

Calculations of the microlensing event rate require integrating the DM density along the line of sight to a source.
In order to greatly simplify their calculation, \citet{LSSTDM} placed all of the expected LSST sources at the centre of the Galactic bulge, so they only needed to calculate the event rate along a single line of sight, and then multiplied by the number of sources to get the final expected number of events. Although this is a reasonable assumption for a crude first estimate, because the amount of dark matter between us and the Galactic bulge is fairly typical of a source star \citep{ML_Bulge_2016MNRAS.463..557W} and many of the sources are in the Galactic bulge, having a single line of sight makes it difficult to understand how sensitive LSST would be to alternative dark object density distributions that may vary over multiple lines of sight. 

Ideally, we would calculate the microlensing event rate to every one of the \revtwo{$\sim 4 \times 10^{9}$ sources that LSST will observe in the r-band with each pass}, but given that each calculation takes around 20 seconds on a single CPU core, this calculation is computationally unfeasible\citep{LSSTScieneBook}. Instead, we simulate a smaller number of LSST sources with a statistically representative distribution of coordinates and distances, calculate the microlensing event rate along each of these lines of sight, and then multiply the sum of microlensing events by the number of real sources per line of sight to get our total number of sources up to $4 \times 10^{9}$.

To create a list of representative LSST sources, we use the software Galaxia \citep{GALAXIA_reference}, which produces simulated surveys based on a complex model of the Milky Way's stellar populations and given survey parameters. We used the LSST \revtwo{single-visit} magnitude limit of \rev{24.5}, using just the i band (\rev{to reflect the superior PSF in the i-band, maximizing the number of resolvable sources}) \citep{Ivezi__2019}. Since microlensing is an achromatic event, \rev{the band chosen for this simulated survey should not have a large impact on our results, as long as the simulated survey approximately captures the distribution of stars resolvable by LSST}. We centred the Galaxia simulated survey on the equatorial south pole, covering an area of 18000 square degrees (to approximately match the observing area of LSST described in \citealt{LSSTDM}). We apply a filter to only return a random fraction of the sources in order to produce a manageable amount of data. Setting this fraction to $f_\mathrm{sample} = 10^{-6}$ \rev{returned a manageable number of sources, while still providing a representative sampling of the Galactic distribution}. The Galactic coordinates of these 86,186 sources are plotted in grey in Fig. \ref{fig:sourcedistribution}, where a clear disk and bulge structure can be seen. \rev{This simulated survey is used as a placeholder for LSST sources, and it would be ideal in future work to use a real source catalogue from LSST once the survey is operational}.

\begin{figure}
    \centering
    \includegraphics[width=0.5\textwidth]{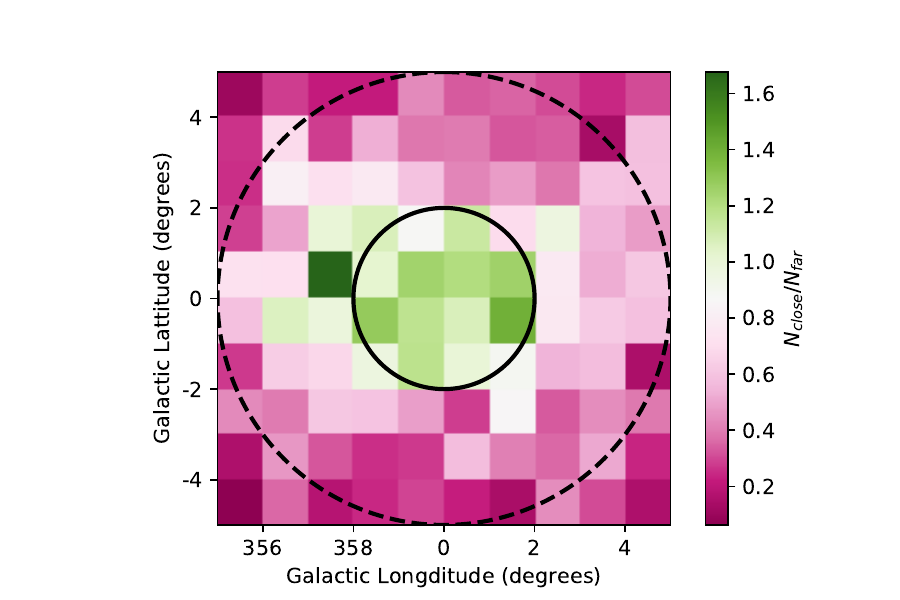}
   \caption{Ratio of the number of simulated sources that would be indistinguishable from their neighbours with an LSST resolution of \rev{0.65 arcseonds} ($N_\mathrm{close}$) compared to the number that are sufficiently far away from their neighbours ($N_\mathrm{far}$) for each 1-pixel square on the sky around the Galactic bulge. Green squares have the majority of their sources indistinguishable from a neighbour, suggesting that LSST will have difficulty measuring individual lightcurves of sources in this region. The circle radii indicate a 2-degree and 5-degree mask of the core, limiting the sources considered to be ones that LSST is likely to be able to resolve.}
    \label{fig:resolvable}
\end{figure}

Even this number of sources is impractically large to perform microlensing calculations on every line of sight. Instead, we randomly select some number of these sources, calculate microlensing rates along those lines of sight, and then multiply the resulting event numbers by an appropriate normalization factor to get the correct total number of sources. The subsamples of sources are shown in various colors in Fig. \ref{fig:sourcedistribution}. 

\rev{One limitation of this simulated survey is that the Galaxia software does not include the Large and Small Magellanic Clouds (LMC and SMC) in their source catalogues. This is unfortunate, as the LMC and SMC will provide some key lines of sight that cut through the Dark Matter Halo while avoiding much of the Galactic plane, therefore probing the spatial lens distribution. Not including these key sightlines makes all of our constraints conservative, as the real LSST source catalogue will include these helpful sightlines. Analyzing the impact of LMC and SMC sightlines of LSST microlensing is left for future work.}

There is some concern that LSST will be unable to resolve closely-spaced sources in dense regions of the sky, such as the Galactic bulge \citep{ML_Bulge_2016MNRAS.463..557W, lsst_resolution}; this will make microlensing with these unresolved sources difficult to analyze. 
In order to estimate how many sources might be unresolved, we calculate the number of simulated Galaxia sources that are close enough to a neighbour that most of the real sources represented by that Galaxia source would be unresolved, based on the estimated resolution of LSST \rev{of 0.65 arcseconds}\citep{LSST_specs}. We use this to calculate where on the sky the majority of sources are likely to be unresolved as opposed to resolvable. Figure \ref{fig:resolvable} shows which 1-degree regions of the sky around the Galactic bulge are expected to have more resolved or unresolved sources \rev{according to our simulated source catalogue}. 
%
%
Since these regions present possible difficulties for microlensing, we create duplicate source catalogues with all sources within either two degrees or five degrees of the Galactic core completely masked from the sample. These masked catalogues provide the opportunity to test the dependence of forecasted constraints on this less reliable bulge region. \revtwo{This masking of of the core will not appreciably change the total number of sources, as LSST is expected to observe around $172$ thousand sources per square degree in that area \citep{LSSTScieneBook}, meaning that a circular mask with radius of five degrees would only mask around $1.4\times 10^7$ sources, which is less than $0.4\%$ of the $4\times10^9$ total sources expected to be resolved each visit. This correction is negligible compared to the uncertainty in the estimated number of LSST sources, and as such is ignored.} \footnote{\revthree{We recognize that there is significant uncertainty on the estimated density of LSST sources in the Galactic core, and that work such as \cite{Schlafly_2018}, \cite{Sajadian_2019}, and \cite{Rich_2020} have reached different estimates than the LSST Science book cited above. However, the key value for our analysis is the total number of sources \textit{outside} the Galactic core, which have fewer uncertainties than the core density. Therefore, we feel comfortable using the internally-consistent values quoted in \cite{LSSTScieneBook}. Real data from LSST will, of course, determine whether these estimated source counts were accurate, and may require future revisions to our estimates.}}

\section{Results: Improved Projected Microlensing Constraints on Primordial Black Holes}
\label{sec:results}

\begin{figure}
    \centering
    \includegraphics[width=0.5\textwidth]{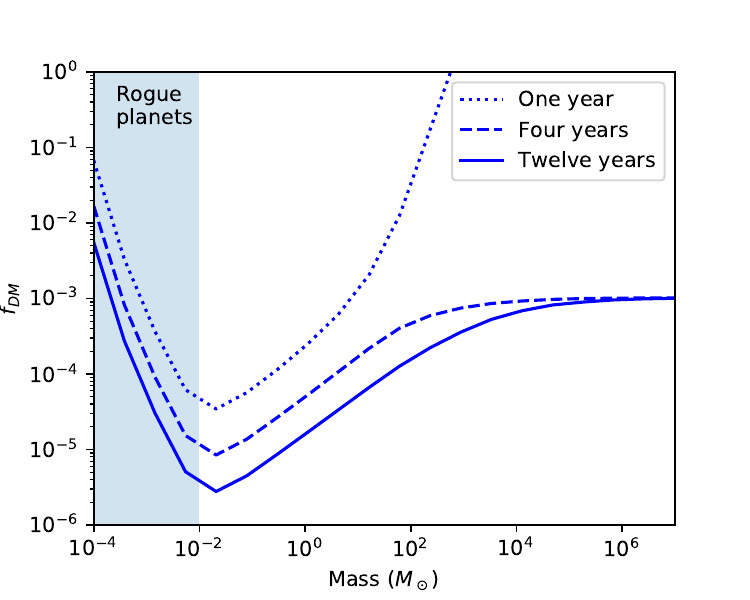}
   \caption{Constraints on the DM fraction of dark objects in a spherical NFW distribution as a function of object mass, with variable survey lengths. All constraints here are calculated using a sampling of the sky distribution with 100 lines of sight. Baryonic foregrounds were not included in this calculation. Depending on the length of time LSST spends surveying the Milky Way, constraints could be significantly improved on low mass objects, as we would expect much higher numbers of these events over the course of the survey. \rev{Note that the high-mass constraints are considerably weaker for a survey just one year in length, as this reduced survey time would make it almost impossible to conduct paralensing with the data, which is essential for constraining high-mass objects.}}
    \label{fig:surveytime}
\end{figure}

\begin{figure}
    \centering
    \includegraphics[width=0.5\textwidth]{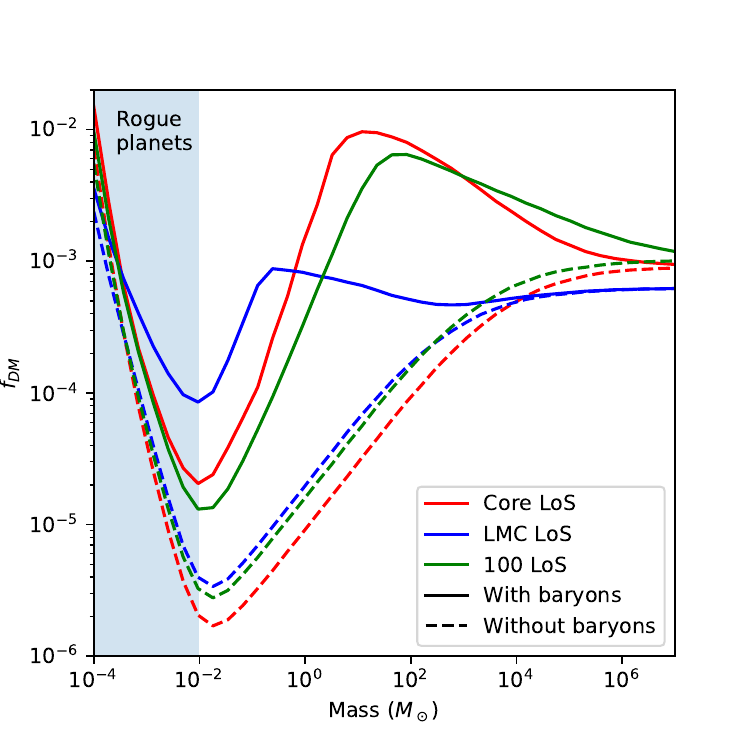}
   \caption{Constraints on the DM fraction of dark objects in a spherical NFW distribution as a function of object mass, showing the effects of multiple lines of sight and incorporating baryonic microlensing. The dashed lines reflect constraints assuming zero baryonic microlensing, while the solid lines represent constraints weakened by the confusion with baryonic microlensing events. With a single line of sight to the Galactic core, the baryon-free constraints are the strongest due to the large amount of DM between us and the core, but adding baryons significantly weakens our core constraints. However, the constraints computed using a more realistic 100 representative lines of sight is less impacted by baryons, as some lines of sight are slightly out of the Galactic disk, and thus less impacted by baryonic microlensing. \rev{Note that the blue curve represents constraints using the same total number of sources as the other constraints, but assuming these sources are all located in the LMC. We understand that LSST will not observe this many sources in the LMC, but we have including these fictional constraints for comparison purposes, 
   to demonstrate the qualitative impact of the LMC sightlines and
   to show the strengths and weaknesses of using a range of sources instead of just the LMC,
   as has been done in past microlensing surveys \citep{Alcock_2000}.}} 
    \label{fig:los_baryons}
\end{figure}

In this section, we explain how each of the changes to the \citet{LSSTDM} methodology mentioned above affected the microlensing rates, and thus changed the constraints on $f_\mathrm{DM}$. The mass-dependent constraints on $f_\mathrm{DM}$ are shown in Fig. \ref{fig:surveytime} and Fig. \ref{fig:los_baryons} for a variety of assumptions, such that we can see the impact of each of the improvements implemented.

The first improvement involved using the microlensing rate integrated over a longer period of time, instead of simply using the optical depth to microlensing at a fixed moment in time. \rev{These results are shown in Figure \ref{fig:surveytime}.} When we use the full LSST length of 12 years, our constraints improve considerably for low masses (shown by the solid line). This is due to the fact that objects of mass $10^{-2} M_\odot$ would have an average crossing time of ~1 day, so we should detect around three orders of magnitude more over the entire length of the twelve-year survey. \rev{This improvement is irrespective of the fractional detectability of these events, as longer survey times will provide more events that could possibly be detected.} Therefore, our constraint projections for objects of this mass improve by around three orders of magnitude. The dashed lines in Fig. \ref{fig:surveytime} shows an intermediate cases of 1 year \rev{or 4 years} spent surveying the Milky Way. 

\begin{figure}
    \centering
    \includegraphics[width=0.5\textwidth]{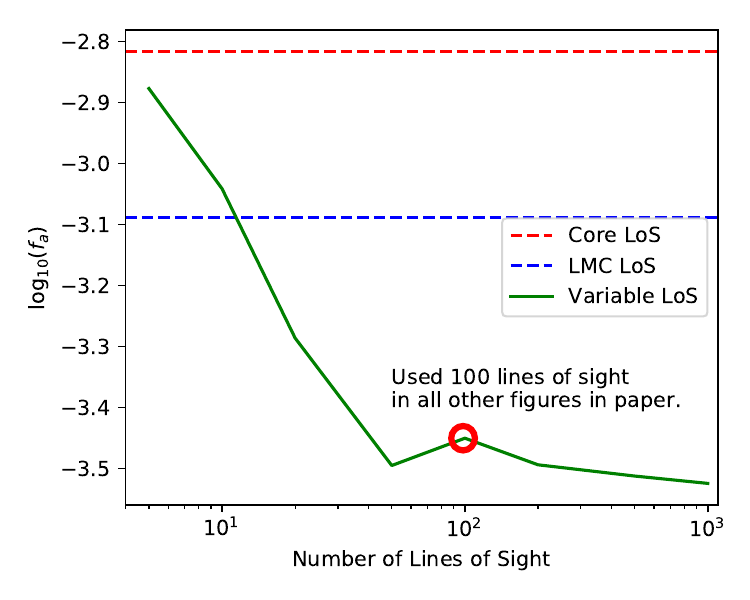}
   \caption{
      \rev{Constraints on the fraction of 1 $M_\odot$ dark objects, as more sight lines representing the sky distribution of LSST sources are added to our analysis. 
      We are assuming a spherical NFW dark object distribution, and including the effects of baryonic backgrounds. 
      It is clear that convergence to the realistic sensitivity is achieved with about  $\sim 100$ lines of sight.
      This also gives  much better constraints than approximating all sightlines as being in either the Galactic core or the LMC. 
      %
      For this reason, we have chosen to use 100 lines of sight for all other plots in this paper, as it is shown here to be accurate to within around 0.1 orders of magnitude, or $30\%$, of the fully converged result.}} 
    \label{fig:sourcenumbers}
\end{figure}

\begin{figure}
    \centering
    \includegraphics[width=0.5\textwidth]{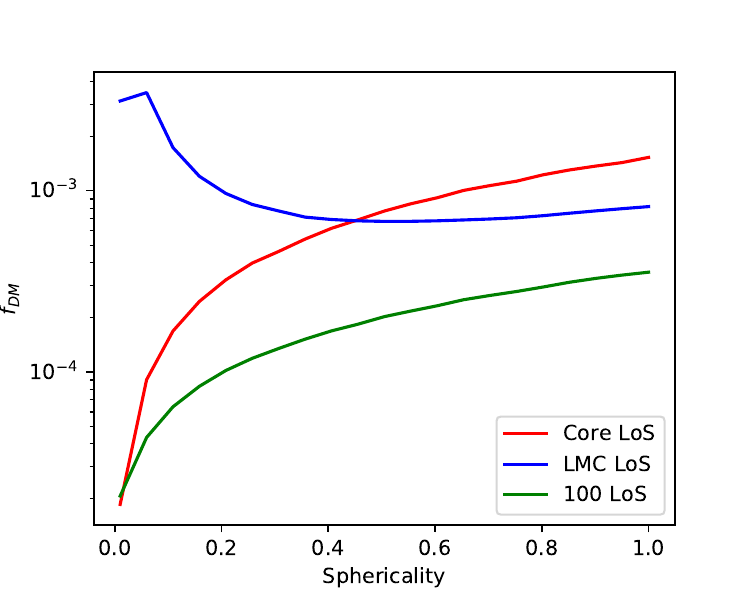}
   \caption{Constraints on the compact DM fraction for one-solar-mass objects as a function of the sphericality ($q$) of the DCO Galactic distribution. The NFW halo is being flattened with respect to parameter $q$ as shown in Equation (10). We can see that using sources in the LMC is less helpful at constraining flattened distributions of DCOs, while approximating all sources to be at the core runs the risk of overestimating the impact of compressed sphericality on the constraints. From this we conclude that using a representative distribution of lines of sight is essential for constraining non-spherical distributions of DCOs.} 
    \label{fig:sphericality}
\end{figure}

 Secondly, we can see in Fig. \ref{fig:los_baryons} the effects of considering baryonic microlensing, and how these baryonic microlensing rates differ for different choices of line of sight. We can see that including baryonic microlensing backgrounds weakens the constraints around the $1 M_\odot$ range, as this is where we would expect many baryonic microlensing events from stars, and thus would require more dark events to warrant a significant detection. This figure also shows how the constraints (and the impact of baryons) changes depending on which lines of sight we use to represent the entire LSST source catalogue in our sensitivity projections. The three colors (red, green, and blue) respond to approximating all the sources as either: living in the Galactic core, living in the Large Magellanic Cloud (LMC; outside the disk and on the far side of the Galaxy), or using 100 random sources drawn from our simulated LSST catalogue respectively \footnote{The LMC is chosen for comparison because it has been used in the past for microlensing surveys such as MACHO \citep{Alcock_2000}. However, placing the LMC constraints alongside those for the core or for a sample of LSST sources is somewhat fictitious, as we assume the same number of total sources located at each of these locations, and it would be impossible to get $4\times10^{9}$ sources in the LMC using existing telescopes. \rev{Also note that LSST may end up using a different observing cadence for the LMC than for the rest of the Galactic plane, and this should be incorporated into any serious analysis of LSST microlensing in the LMC.}}. 
 
 The slight mass-shift of the baryonic bump in the three cases is \rev{primarily} the result of different perpendicular velocities of baryonic stars along each of the three lines of sight. The core constraints (in red) are made in a region with a lot of baryonic stars (along the plane of the Galactic disk), and as a result the bump there is quite large. The LMC constraints, on the other hand, have a line of sight that passes through the far side of the Galactic disk as well, so it will encounter stars moving in the opposite direction to Earth's orbit. This will lead to the baryonic microlensing events being shorter in duration, so they would be confused for lower-mass dark microlensing events, which is why the baryon-bump for the LMC source is shifted to the left. For the 100 LSST lines of sight, some lines of sight will be looking along the disk either to the left or right of the Galactic core, resulting in the transverse velocities of the baryonic stars being much smaller than when looking directly into the core. This makes those microlensing events longer in duration, so they will be confused for more massive dark objects. \rev{We also note that the distribution of objects along the line of sight (ie. the distance) can also impact the crossing times, and thus which mass ranges are most impacted by baryonic microlensing. However, since the distance distribution of dark and baryonic objects for most lines of sight is qualitatively similar (increasing towards the Galactic centre, and decreasing away from it), we conclude that changes in the transverse velocities are the driving reason behind the mass-shifts of the bumps seen in Figure \ref{fig:los_baryons}.} This result illustrates the importance of properly incorporating the effects of different lines of sight and velocities on baryonic microlensing.

Lastly, we can characterize the impact of including multiple lines of sight, as opposed to calculating the microlensing rate for a single line of sight between us and the Galactic bulge. We can see in Fig. \ref{fig:los_baryons} that using 100 lines of sight as opposed to one core line of sight changes the impact of baryonic microlensing rates over a range of object masses. Fig. \ref{fig:sourcenumbers} shows how the constraints for one solar-mass dark objects change depending on the number of lines of sight. For reference, the red line shows the constraints for a single line of sight between us and the Galactic bulge, while the blue line is a reference for constraints using the LMC. We can see that the constraints change as we use more lines of sight, but begin to stabilize after around 50 lines of sight. We use 100 lines of sight for all other calculations in this paper, as it is within 0.1 orders of magnitude of the limit at 1000 lines of sight, but is still able to be computed in a reasonable amount of time. Our final projected sensitivity constraints can therefore be interpreted as having a $\sim 30\%$ level of statistical error, which is sufficient for our purposes.

\section{Results: Projected Constraints on Alternative Density Models}
\label{sec:alt_dens}
In the previous section, we explained how our estimates of microlensing constraints improve upon previous calculations, and how they are sensitive to events at many locations in the Galaxy. These modifications produce a more realistic forecast of constraints on regular spherically-distributed models of compact DM, as shown in Fig. \ref{fig:old_constraints}.
Another important advantage of these improvements is that they allow us to reliably predict the sensitivity of LSST for non-spherical distributions of compact DM, as was motivated in the Introduction. For example, one interesting case is that of a mirror disk of dark compact objects, resembling our baryonic Galactic disk up to some rescaling or rotation, and comprising only a few percent of the total dark matter in our Galaxy. Note that only a fraction of the total mirror matter in the dark disk would be expected to have formed compact objects, just as in the visible disk, and it is this compact dark object density distribution that we are concerned with. Their low fraction of the total DM abundance makes it difficult for these dark disks to be constrained by Galactic dynamics, especially if they have non-negligible disk height (for constraints on a very ``thin'' disk scenario from stellar kinematics, see \cite{Schutz:2017tfp} and \cite{Buch:2018qdr}).
Fortunately, microlensing has the potential to set tight constraints on the compact-object fraction for arbitrary Galactic distributions.

\begin{figure*}
   \noindent
    \hspace*{-2.5cm}\includegraphics[width=1.2\textwidth]{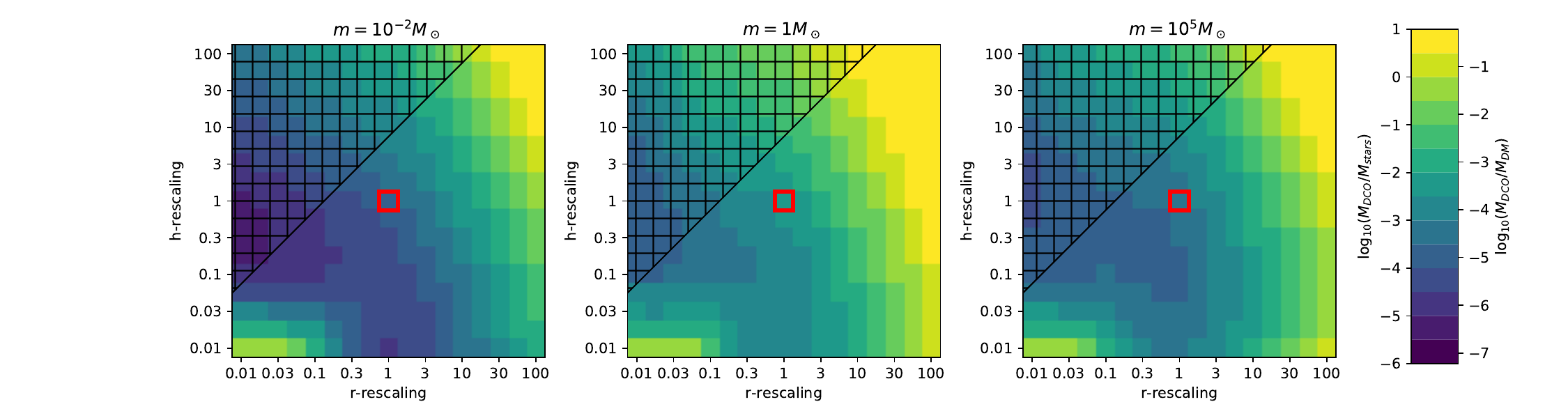}
   \caption{Constraints on the ratio of the total mass of dark compact objects (DCOs) to ordinary stars, depending on how the vertical and radial length scales of the dark disk compare to those of the Milky Way (see Eq. (11)). The color indicates maximum allowed DM mass, either as a fraction of total baryonic matter or a fraction of total dark matter (hence the two number scales on the colorbar). The red square indicates a perfectly symmetrical mirror disk (with the same dimensions as the Milky Way's baryonic disk), and we can see that the constraints get tighter when the disk is compressed. However, this improvement soon reaches a saturation, as further compression (beyond around 0.3 in radial compression, and 0.1 in vertical compression) weakens the constraints. Note that these bounds only apply quantitatively to sharply peaked mass distributions for the DCOs. The black cross-hatched region corresponds to `disks' that are taller than they are wide, which is unphysical, but we still show the corresponding results as a demonstration of the versatility of our forecasting software. }
    \label{fig:rescaled}
\end{figure*}

\begin{figure}
    \includegraphics[width=0.5\textwidth]{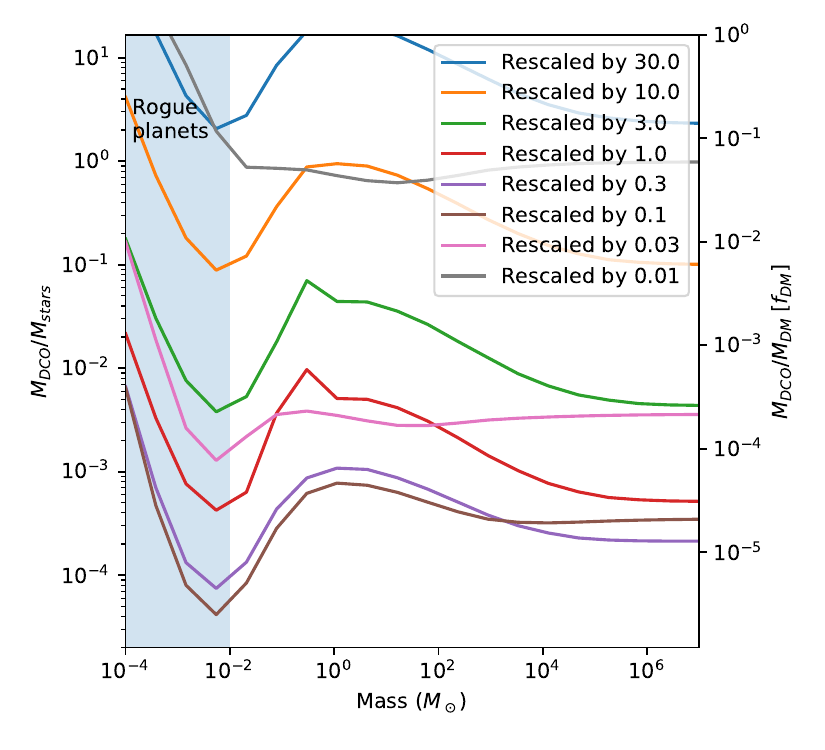}
   \caption{Forecast constraints on the total mass of DCOs in a dark disk distribution as a function of object mass. The lines show constraints for various spatial rescaling factors, where the radial and vertical rescaling with respect to the baryonic disk are identical (see Eq. (11)). The total mass of the dark disk is kept fixed during this rescaling, so the smaller dark disk has much more mass concentrated near the centre of the Galaxy, corresponding to tighter constraints on the total allowed mass. Note that these bounds only apply quantitatively to sharply peaked mass distributions for the DCOs. The left axis shows the constraints on dark compact object mass as a fraction of total stellar mass in the Milky Way, while the right axis shows the constraints as a fraction of total dark matter in the Galaxy (in other words, $f_\mathrm{DM}$). As was seen in Fig. \ref{fig:rescaled}, the improvements in constraints for compact disks reaches a saturation point at around 0.1 before it begins to weaken again. We can also see that the role of baryonic backgrounds becomes slightly less important for highly compact disks, as seen by the reduced `bump' in the pink and grey curves. The constraints for disks rescaled by a factor of 100 are not shown on this plot, as they lie completely above the top of the figure, and cannot be constrained to be anything less than 100\% of the DM for any mass range.} 
    \label{fig:scale_mass}
\end{figure}

\subsection{Flattened NFW halo}
\label{sec:ad_flat_halo}

In principle, any Galactic density function could be used for the dark compact objects in our extended calculation. To get a basic understanding of the role that flatness plays in our compact DM constraints, we first consider a simple `squashed' NFW distribution.
This connects straightforwardly to our previous constraints on a spherical NFW distribution. For all previous tests, we were using a spherical NFW distribution of equation \eqref{eq-nfw}. To convert this profile into a squashed NFW distribution, we convert the above spherically-symmetric distribution to cylindrical coordinates, and rescale the $z$-axis by some factor $q$, being sure to add a normalization factor to maintain constant total mass:
\begin{equation}
\rho_{\mathrm{NFW};q}(r,z) = \frac{\rho_0}{q \big(\frac{\sqrt{r^2 + (z/q)^2}}{R_s}\big)\big( 1 + \frac{\sqrt{r^2 + (z/q)^2}}{R_s}\big)^2}.
\end{equation}
We can think of this new parameter $q$ as the `sphericality', where $q=1$ corresponds to a normal spherical NFW distribution, and $q=0$ corresponds to an infinitely thin delta-function disk. We can then re-run our microlensing constraints with a range of sphericalities, and for a variety of source locations. The results are shown in Fig. \ref{fig:sphericality}. We can see that the constraints on the compact dark object fraction get tighter for flatter distributions. This is not unexpected, as more of the dark objects are located along the lines of sight to stars in the Galactic disk. We can also see that the dependence on sphericality changes for different collections of lines of sight. This reinforces the importance of using a representative sample of sightlines when investigating non-standard distributions of dark compact objects, and not simply approximating all sources to be at the Galactic core.

\subsection{Rescaled mirror disk}
\label{sec:ad_mirror_disk}

A more realistic model of aspherically-distributed dark compact objects would be a dark disk with a similar distribution to our own. Therefore, it is interesting to ask what constraints we could put on a dark object distribution resembling the double-exponential model used in our baryonic microlensing. We can try rescaling this model disk in both the vertical and radial directions (renormalizing to maintain a given total mass), and observe how our constraints on the total mass ratio between mirror and baryonic matter changes. This new density formula for a dark disk would have the form
\begin{equation}
    \rho_\mathrm{DD; q_r, q_z}(r,z) = \frac{A_{DD}}{q_r^2 q_z} \exp\bigg(-\frac{(r/q_r - R_0)}{h_R}\bigg) \exp\bigg(-\frac{|z/q_z|}{h_z}\bigg)
\end{equation}
with $A_{DD}$ being the amplitude of the dark disk distribution, and $q_R$ and $q_z$ being the radial and vertical rescaling factors, respectively. We use 100 lines of sight to calculate the constraints on the rescaled dark disk. Fig. \ref{fig:rescaled} shows the projected constraints on dark compact objects of mass $10^{-2} M_\odot, 1 M_\odot$, and $10^\rev{5} M_\odot$  as a function of vertical and radial rescaling, while Fig. \ref{fig:scale_mass} highlights how these constraints depend on the mass of dark compact objects, computed for dark disks rescaled equally in the horizontal and vertical directions. Note that although the Galactic distribution of the dark and baryonic objects is similar, the mass distributions are nonetheless different (with dark objects having a delta-function mass distribution, while baryonic stars have a Kroupa IMF as described above). A more thorough investigation of dark object mass distributions is left for future work.

 These results show that constraints on dark compact objects with a disk distribution can get significantly tighter if the disk is concentrated towards the centre of the Galaxy. 
 This is reasonable, as the density along some lines of sight increases as the disk is compacted. However, it is interesting to note that this is only true down to a rescaling of around 0.3 -- beyond this point the constraints get weaker again. This is likely due to the fact that highly compressed dark disks only overlap with a small number of sources, so the increase in statistical power for these central lines of sight is offset by a reduction in some of the more peripheral lines of sight. 
 %

\subsection{Tilted mirror disk}
\label{sec:ad_tilt}

\begin{figure}
    \includegraphics[width=0.5\textwidth]{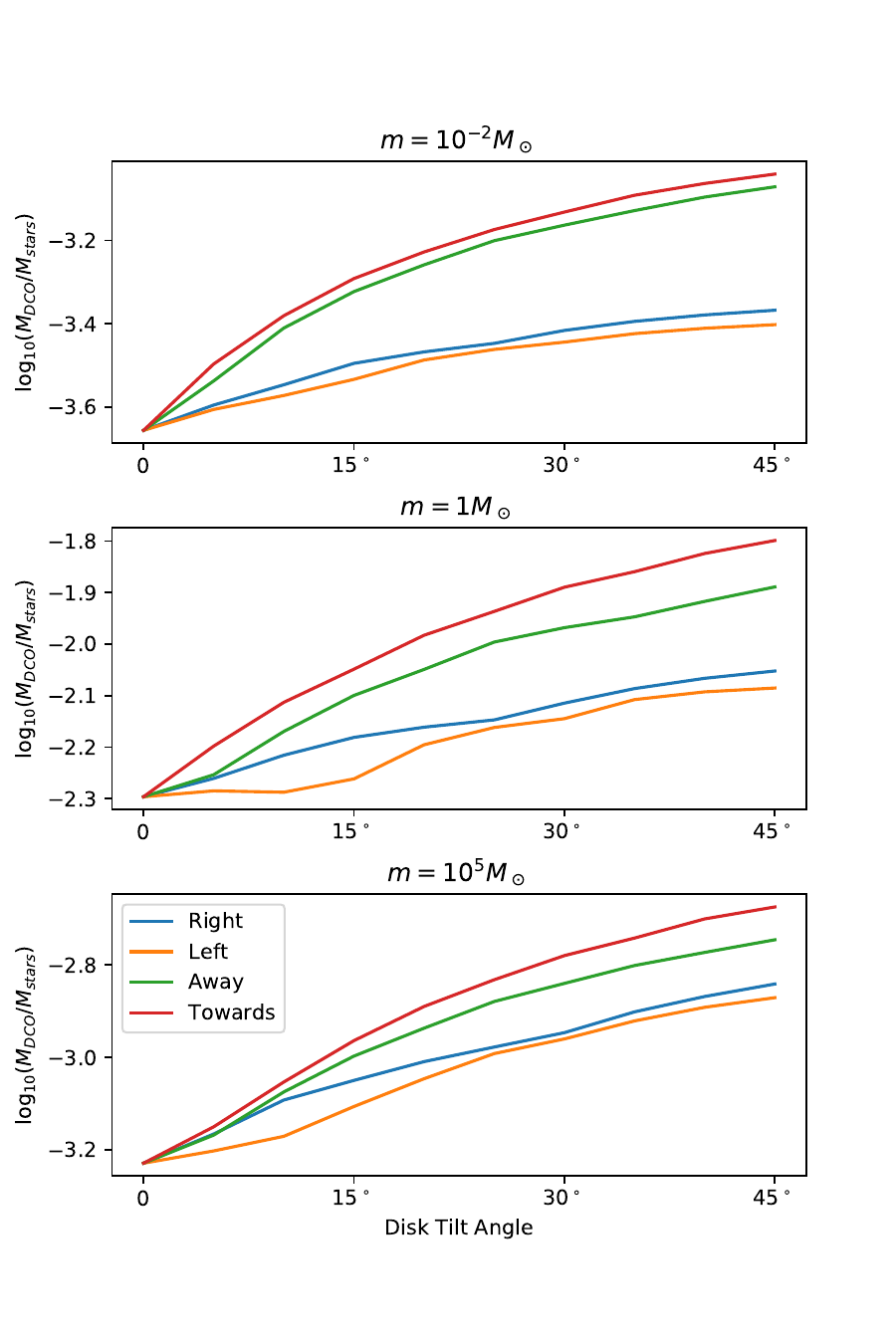}
   \caption{Forecasted constraints on the ratio of the total masses of dark compact objects (DCOs) in a dark disk and ordinary stars, depending on the tilt of the dark disk relative to our own. The four lines illustrate the result of tilting the dark disk in different directions, as viewed from Earth looking towards the Galactic centre. We can see that the constraints get weaker when we tilt the dark disk, regardless of the direction of tilt. This plot assumes a dark disk with the same dimensions as the Milky Way's baryonic disk, and each subplot shows constraints for a different mass of dark compact object. We use 100 lines of sight in order to achieve a representative sample of the dark object distribution.} 
    \label{fig:tilted}
\end{figure}

Another possible distribution of mirror matter that could be considered is a dark disk that is tilted with respect to our own disk. This could be caused by imperfect coupling between the dark and baryonic sectors as our Galactic disks were forming, resulting in imperfect alignment between the angular momentum vectors (although gravitationally they should eventually align). We can see in Fig.~\ref{fig:tilted} how the constraints on the disk mass ratios get weaker with a tilted disk, likely due to the fact that we have fewer sources outside the Galactic plane, and thus less statistical power. These constraints are for $10^{-2} M_\odot, 1 M_\odot$, and $10^5 M_\odot$ objects, using 100 lines of sight. 

It is interesting to note that although the constraints get weaker when the dark disk is tilted in any direction, there is a consistent difference in how much weaker depending on the tilt angle. Tilting the disk to the left or right (from the perspective of our Sun looking into the Galactic core) preserves some of the key lines of sight through the dark disk into the Galactic bulge (where many sources are located), so the constraints only get slightly weaker. Tilting towards or away from us, on the other hand, results in no good lines of sight that go along the plane of the dark disk, so they get significantly weaker. The differences between left and right, as well as the differences between towards and away, could be the result of rotation of the dark and baryonic disks, or asymmetries in the region of the sky seen by LSST (as seen in Fig.~\ref{fig:sourcedistribution}). 

Finally, we point out that it is unlikely that a dark disk would exhibit a tilting angle as high as the maximum of $45^\circ$ that we show in our plot, but our calculation demonstrates that more reasonable tilting angles might be detectable, even if their impact on microlensing sensitivity is quite modest. 

\subsection{Future Applications of Microlensing Code}
\label{sec:fut_app}


In addition to the spherical and flattened NFW profile, and the rescaled and tilted double-exponential disk, our code for calculating these microlensing constraints could be used on any conceivable distribution of dark compact objects, corresponding to more exotic theories of compact DM, or more precise future predictions for the distribution of dark compact objects in a given theory of dissipative DM. These investigations could be performed very easily by supplying the code with a new functional form of the Galactic spatial distribution of DCOs.
A simple example would be to investigate dark disks that are simultaneously rescaled and tilted, which was not done in this work. Another possibility would be to consider distributions of dark compact objects that mirror some other component of our baryonic Galaxy, perhaps only forming a `dark bulge' or `dark bar'. More complex models could include dark compact objects that are clustered, similar to baryonic globular clusters. These dark object distributions could also be informed from specific models of compact dark matter, such as dissipative dark matter, axion stars, and mirror dark matter. This would require theoretical predictions for the Galactic distribution of these specific models of dark compact objects, perhaps dependent on the free parameters of the underlying model, which has yet to be done for most of the models of compact DM mentioned here. In these cases of theoretically-motivated Galactic distributions, our constraints on the parameters of the distribution function could translate into interesting constraints on the parameters of the underlying model. 

An only slightly more-involved extension of our analysis that we have not yet investigated is constraints on more general distributions of dark object masses, instead limiting our investigations to delta-function mass distributions of DCOs. As a first approximation, one might simply convolve the mass distribution with the mass-dependent constraints on dark object fraction that the code currently produces. However, this would fail to accurately capture the effect of baryonic ``background events" in crossing time.
The correct approach would be to compare the distributions of dark and baryonic events as a function of crossing time.
This would require some modification to the code, in order to compute dark event rates for a range of masses, and then weighting those distributions of events by some mass function to get a total distribution of events in crossing time. Comparing this to the distribution of baryonic microlensing events could produce interesting effects. For example, when we restricted our baryonic star mass distribution to a delta function, we found that our constraints got significantly weaker when the masses overlapped perfectly. Similarly, we expect that a dark object mass distribution perfectly mirroring a stellar IMF would have significantly weaker constraints, as the event distribution would perfectly match the baryonic events. In this case, the constraints would be limited only by the inherent uncertainty on baryonic star abundance, which we currently assume to be around 5\% (consistent with \citealt{stardense_Bovy_2017}). Realistic sensitivities likely lie between these two extremes, depending on how much a given DCO mass distribution ``stands out'' against the background of the baryonic stellar mass distribution. We leave a more thorough investigation of the impact of DCO mass distribution to future work.

\section{Conclusion}
\label{sec:conclusion}

LSST is expected to produce some of the strongest microlensing constraints on dark compact objects in our Galaxy. As this paper has shown, LSST will not just be able to constrain models of PBHs, but it has the potential to constrain many other more exotic models of dark compact objects that could make up a fraction of the DM in our Galaxy. 

Of special interest is the potential to constrain models of dissipative dark matter, such as the Mirror Twin Higgs and related models. These models predict that a subcomponent of dark matter could be in the form of mirror baryons and electrons, forming atomic dark matter that can cool and collapse into mirror stars. These compact objects can have unusual Galactic distributions, which motivates more complex microlensing constraint calculations. This paper has improved on older LSST microlensing constraints by including variable observing times, baryonic microlensing foregrounds, and multiple lines of sight characteristic of LSST sources. 

Using these improvements, we are not only able to improve sensitivity projections for primordial black holes, but also forecast novel constraints on a variety of alternative DM distributions. Some example models that were constrained include a squashed NFW distribution, a rescaled mirror disk, and a tilted mirror disk. For example, we forecast that LSST will be able to constrain dark compact objects with one solar mass in an NFW distribution to have a DM fraction under $4.1\times10^{-4}$. One-solar-mass objects with a delta function mass distribution and a dark disk spatial distribution with the same dimensions as the baryonic Galactic disk will be constrained to below $f_{DM} < 3.1\times10^{-4}$, while those with masses of $10^5 M_{\odot}$ will be constrained to below $f_{DM} < 3.4\times10^{-5}$ %
This represents a significant advance in our understanding of microlensing probes of non-minimal dark matter, and suggests that realistic DCO distributions could be detected with percent or permille dark matter fractions in our Milky Way disk. Our results motivate further work to understand the fraction of mirror dark matter that actually forms DCOs, as well as their mass distributions, which will have a major impact on the interpretation of these bounds in terms of more fundamental particle physics theory parameters.
The software created for this paper can easily be re-purposed to constrain any arbitrary spatial distribution of compact objects. All the code used to produce the figures in this paper, as well as compute new constraints on custom Galactic distributions, is available at \href{https://github.com/HarrisonWinch96/DarkDisk_Microlensing}{\textcolor{blue}{this GitHub link}}~\citep{harrisongithub}.

\section{Acknowledgements}
\label{sec:acknowledgements}

We would like to thank Ren\'ee Hlo\v zek, Will Dawson, David Hendel, and Nathaniel Starkman for their technical guidance and helpful discussions.

HW would like to acknowledge the support of the Natural Sciences and Engineering Research Council of Canada (NSERC) Canadian Graduate Scholarships - Master's Program, [funding reference number 542579 / 2019].
JB acknowledges financial support from NSERC (funding reference numbers RGPIN-2015-05235 \& RGPIN-2020-04712), an Ontario Early Researcher Award (ER16-12-061), and from the Canada Research Chair program.
 The research of DC and JS was supported by a Discovery Grant from the Natural Sciences and Engineering Research Council of Canada, and by the Canada Research Chair program.

\bibliographystyle{aasjournal}
\bibliography{main}

\begin{thebibliography}{}
\expandafter\ifx\csname natexlab\endcsname\relax\def\natexlab#1{#1}\fi
\providecommand{\url}[1]{\href{#1}{#1}}
\providecommand{\dodoi}[1]{doi:~\href{http://doi.org/#1}{\nolinkurl{#1}}}
\providecommand{\doeprint}[1]{\href{http://ascl.net/#1}{\nolinkurl{http://ascl.net/#1}}}
\providecommand{\doarXiv}[1]{\href{https://arxiv.org/abs/#1}{\nolinkurl{https://arxiv.org/abs/#1}}}

\bibitem[{Abbott {et~al.}(2020)Abbott, Abbott, Abraham, Acernese, Ackley,
  Adams, Adhikari, Adya, Affeldt, Agathos, Agatsuma, Aggarwal, Aguiar, Aich,
  Aiello, Ain, Ajith, Akcay, Allen, Allocca, Altin, Amato, Anand, Ananyeva,
  Anderson, Anderson, Angelova, Ansoldi, Antier, Appert, Arai, Araya, Areeda,
  Ar\`ene, Arnaud, Aronson, Arun, Asali, Ascenzi, Ashton, Aston, Astone, Aubin,
  Aufmuth, AultONeal, Austin, Avendano, Babak, Bacon, Badaracco, Bader, Bae,
  Baer, Baird, Baldaccini, Ballardin, Ballmer, Bals, Balsamo, Baltus, Banagiri,
  Bankar, Bankar, Barayoga, Barbieri, Barish, Barker, Barkett, Barneo, Barone,
  Barr, Barsotti, Barsuglia, Barta, Bartlett, Bartos, Bassiri, Basti, Bawaj,
  Bayley, Bazzan, B\'ecsy, Bejger, Belahcene, Bell, Beniwal, Benjamin, Bentley,
  Bergamin, Berger, Bergmann, Bernuzzi, Berry, Bersanetti, Bertolini,
  Betzwieser, Bhandare, Bhandari, Bidler, Biggs, Bilenko, Billingsley, Birney,
  Birnholtz, Biscans, Bischi, Biscoveanu, Bisht, Bissenbayeva, Bitossi,
  Bizouard, Blackburn, Blackman, Blair, Blair, Blair, Bobba, Bode, Boer,
  Boetzel, Bogaert, Bondu, Bonilla, Bonnand, Booker, Boom, Bork, Boschi, Bose,
  Bossilkov, Bosveld, Bouffanais, Bozzi, Bradaschia, Brady, Bramley, Branchesi,
  Brau, Breschi, Briant, Briggs, Brighenti, Brillet, Brinkmann, Brockill,
  Brooks, Brooks, Brown, Brunett, Bruno, Bruntz, Buikema, Bulik, Bulten,
  Buonanno, Buscicchio, Buskulic, Byer, Cabero, Cadonati, Cagnoli, Cahillane,
  Calder\'on~Bustillo, Callaghan, Callister, Calloni, Camp, Canepa, Cannon,
  Cao, Cao, Carapella, Carbognani, Caride, Carney, Carullo, Casanueva~Diaz,
  Casentini, Casta\~neda, Caudill, Cavagli\`a, Cavalier, Cavalieri, Cella,
  Cerd\'a-Dur\'an, Cesarini, Chaibi, Chakravarti, Chan, Chan, Chandra, Chao,
  Charlton, Chase, Chassande-Mottin, Chatterjee, Chaturvedi, Chatziioannou,
  Chen, Chen, Chen, Cheng, Cheong, Chia, Chiadini, Chierici, Chincarini,
  Chiummo, Cho, Cho, Cho, Christensen, Chu, Chua, Chung, Chung, Ciani,
  Ciecielag, Cie\ifmmode~\acute{s}\else \'{s}\fi{}lar, Ciobanu, Ciolfi,
  Cipriano, Cirone, Clara, Clark, Clearwater, Clesse, Cleva, Coccia, Cohadon,
  Cohen, Colleoni, Collette, Collins, Colpi, Constancio, Conti, Cooper, Corban,
  Corbitt, Cordero-Carri\'on, Corezzi, Corley, Cornish, Corre, Corsi, Cortese,
  Costa, Cotesta, Coughlin, Coughlin, Coulon, Countryman, Couvares, Covas,
  Coward, Cowart, Coyne, Coyne, Creighton, Creighton, Cripe, Croquette,
  Crowder, Cudell, Cullen, Cumming, Cummings, Cunningham, Cuoco, Curylo,
  Canton, D\'alya, Dana, Daneshgaran-Bajastani, D'Angelo, Danilishin,
  D'Antonio, Danzmann, Darsow-Fromm, Dasgupta, Datrier, Dattilo, Dave, Davier,
  Davies, Davis, Daw, DeBra, Deenadayalan, Degallaix, De~Laurentis,
  Del\'eglise, Delfavero, De~Lillo, Del~Pozzo, DeMarchi, D'Emilio, Demos, Dent,
  De~Pietri, De~Rosa, De~Rossi, DeSalvo, de~Varona, Dhurandhar, D\'{\i}az,
  Diaz-Ortiz, Dietrich, Di~Fiore, Di~Fronzo, Di~Giorgio, Di~Giovanni,
  Di~Giovanni, Di~Girolamo, Di~Lieto, Ding, Di~Pace, Di~Palma, Di~Renzo,
  Divakarla, Dmitriev, Doctor, Donovan, Dooley, Doravari, Dorrington, Downes,
  Drago, Driggers, Du, Ducoin, Dupej, Durante, D'Urso, Dwyer, Easter, Eddolls,
  Edelman, Edo, Edy, Effler, Ehrens, Eichholz, Eikenberry, Eisenmann,
  Eisenstein, Ejlli, Errico, Essick, Estelles, Estevez, Etienne, Etzel, Evans,
  Evans, Ewing, Fafone, Fairhurst, Fan, Farinon, Farr, Farr, Fauchon-Jones,
  Favata, Fays, Fazio, Feicht, Fejer, Feng, Fenyvesi, Ferguson,
  Fernandez-Galiana, Ferrante, Ferreira, Ferreira, Fidecaro, Fiori, Fiorucci,
  Fishbach, Fisher, Fittipaldi, Fitz-Axen, Fiumara, Flaminio, Floden, Flynn,
  Fong, Font, Forsyth, Fournier, Frasca, Frasconi, Frei, Freise, Frey, Frey,
  Fritschel, Frolov, Fronz\`e, Fulda, Fyffe, Gabbard, Gadre, Gaebel, Gair,
  Galaudage, Ganapathy, Ganguly, Gaonkar, Garc\'{\i}a-Quir\'os, Garufi,
  Gateley, Gaudio, Gayathri, Gemme, Genin, Gennai, George, George, Gergely,
  Ghonge, Ghosh, Ghosh, Ghosh, Giacomazzo, Giaime, Giardina, Gibson, Gier,
  Gill, Glanzer, Gniesmer, Godwin, Goetz, Goetz, Gohlke, Goncharov, Gonz\'alez,
  Gopakumar, Gossan, Gosselin, Gouaty, Grace, Grado, Granata, Grant, Gras,
  Grassia, Gray, Gray, Greco, Green, Green, Gretarsson, Griggs, Grignani,
  Grimaldi, Grimm, Grote, Grunewald, Gruning, Guidi, Guimaraes, Guix\'e,
  Gulati, Guo, Gupta, Gupta, Gupta, Gustafson, Gustafson, Haegel, Halim, Hall,
  Hamilton, Hammond, Haney, Hanke, Hanks, Hanna, Hannam, Hannuksela, Hansen,
  Hanson, Harder, Hardwick, Haris, Harms, Harry, Harry, Hasskew, Haster,
  Haughian, Hayes, Healy, Heidmann, Heintze, Heinze, Heitmann, Hellman, Hello,
  Hemming, Hendry, Heng, Hennes, Hennig, Heurs, Hild, Hinderer, Hoback,
  Hochheim, Hofgard, Hofman, Holgado, Holland, Holt, Holz, Hopkins, Horst,
  Hough, Howell, Hoy, Huang, H\"ubner, Huerta, Huet, Hughey, Hui, Husa,
  Huttner, Huxford, Huynh-Dinh, Idzkowski, Iess, Inchauspe, Ingram, Intini,
  Isac, Isi, Iyer, Jacqmin, Jadhav, Jadhav, James, Jani, Janthalur, Jaranowski,
  Jariwala, Jaume, Jenkins, Jiang, Johns, Johnson-McDaniel, Jones, Jones,
  Jones, Jones, Jones, Jonker, Ju, Junker, Kalaghatgi, Kalogera, Kamai,
  Kandhasamy, Kang, Kanner, Kapadia, Karki, Kashyap, Kasprzack, Kastaun,
  Katsanevas, Katsavounidis, Katzman, Kaufer, Kawabe, K\'ef\'elian, Keitel,
  Keivani, Kennedy, Key, Khadka, Khalili, Khan, Khan, Khan, Khazanov, Khetan,
  Khursheed, Kijbunchoo, Kim, Kim, Kim, Kim, Kim, Kim, Kim, Kimball, King,
  Kinley-Hanlon, Kirchhoff, Kissel, Kleybolte, Klimenko, Knowles, Knyazev,
  Koch, Koehlenbeck, Koekoek, Koley, Kondrashov, Kontos, Koper, Korobko, Korth,
  Kovalam, Kozak, Kringel, Krishnendu, Kr\'olak, Krupinski, Kuehn, Kumar,
  Kumar, Kumar, Kumar, Kumar, Kuo, Kutynia, Lackey, Laghi, Lalande, Lam,
  Lamberts, Landry, Lane, Lang, Lange, Lantz, Lanza, La~Rosa, Lartaux-Vollard,
  Lasky, Laxen, Lazzarini, Lazzaro, Leaci, Leavey, Lecoeuche, Lee, Lee, Lee,
  Lee, Lee, Lehmann, Leroy, Letendre, Levin, Li, Li, li, Li, Li, Linde, Linker,
  Linley, Littenberg, Liu, Liu, Llorens-Monteagudo, Lo, Lockwood, London,
  Longo, Lorenzini, Loriette, Lormand, Losurdo, Lough, Lousto, Lovelace,
  L\"uck, Lumaca, Lundgren, Ma, Macas, Macfoy, MacInnis, Macleod, MacMillan,
  Macquet, Maga\~na Hernandez, Maga\~na Sandoval, Magee, Majorana, Maksimovic,
  Malik, Man, Mandic, Mangano, Mansell, Manske, Mantovani, Mapelli, Marchesoni,
  Marion, M\'arka, M\'arka, Markakis, Markosyan, Markowitz, Maros, Marquina,
  Marsat, Martelli, Martin, Martin, Martinez, Martynov, Masalehdan, Mason,
  Massera, Masserot, Massinger, Masso-Reid, Mastrogiovanni, Matas, Matichard,
  Mavalvala, Maynard, McCann, McCarthy, McClelland, McCormick, McCuller,
  McGuire, McIsaac, McIver, McManus, McRae, McWilliams, Meacher, Meadors,
  Mehmet, Mehta, Mejuto~Villa, Melatos, Mendell, Mercer, Mereni, Merfeld,
  Merilh, Merritt, Merzougui, Meshkov, Messenger, Messick, Metzdorff, Meyers,
  Meylahn, Mhaske, Miani, Miao, Michaloliakos, Michel, Middleton, Milano,
  Miller, Millhouse, Mills, Milotti, Milovich-Goff, Minazzoli, Minenkov,
  Mishkin, Mishra, Mistry, Mitra, Mitrofanov, Mitselmakher, Mittleman, Mo,
  Mogushi, Mohapatra, Mohite, Molina-Ruiz, Mondin, Montani, Moore, Moraru,
  Morawski, Moreno, Morisaki, Mours, Mow-Lowry, Mozzon, Muciaccia, Mukherjee,
  Mukherjee, Mukherjee, Mukherjee, Mukund, Mullavey, Munch, Mu\~niz, Murray,
  Nagar, Nardecchia, Naticchioni, Nayak, Neil, Neilson, Nelemans, Nelson, Nery,
  Neunzert, Ng, Ng, Nguyen, Nguyen, Nichols, Nichols, Nissanke, Nitz, Nocera,
  Noh, North, Nothard, Nuttall, Oberling, O'Brien, Oganesyan, Ogin, Oh, Oh,
  Ohme, Ohta, Okada, Oliver, Olivetto, Oppermann, Oram, O'Reilly, Ormiston,
  Ortega, O'Shaughnessy, Ossokine, Osthelder, Ottaway, Overmier, Owen, Pace,
  Pagano, Page, Pagliaroli, Pai, Pai, Palamos, Palashov, Palomba, Pan, Panda,
  Pang, Pankow, Pannarale, Pant, Paoletti, Paoli, Parida, Parker, Pascucci,
  Pasqualetti, Passaquieti, Passuello, Patricelli, Payne, Pearlstone, Pechsiri,
  Pedersen, Pedraza, Pele, Penn, Perego, Perez, P\'erigois, Perreca, Perri\`es,
  Petermann, Pfeiffer, Phelps, Phukon, Piccinni, Pichot, Piendibene,
  Piergiovanni, Pierro, Pillant, Pinard, Pinto, Piotrzkowski, Pirello, Pitkin,
  Plastino, Poggiani, Pong, Ponrathnam, Popolizio, Porter, Powell, Prajapati,
  Prasai, Prasanna, Pratten, Prestegard, Principe, Prodi, Prokhorov, Punturo,
  Puppo, P\"urrer, Qi, Quetschke, Quinonez, Raab, Raaijmakers, Radkins,
  Radulesco, Raffai, Rafferty, Raja, Rajan, Rajbhandari, Rakhmanov, Ramirez,
  Ramos-Buades, Rana, Rao, Rapagnani, Raymond, Razzano, Read, Regimbau, Rei,
  Reid, Reitze, Rettegno, Ricci, Richardson, Richardson, Ricker,
  Riemenschneider, Riles, Rizzo, Robertson, Robinet, Rocchi, Rodriguez-Soto,
  Rolland, Rollins, Roma, Romanelli, Romano, Romel, Romero-Shaw, Romie, Rose,
  Rose, Rose, Rosi\ifmmode~\acute{n}\else \'{n}\fi{}ska, Rosofsky, Ross, Rowan,
  Rowlinson, Roy, Roy, Roy, Ruggi, Rutins, Ryan, Sachdev, Sadecki,
  Sakellariadou, Salafia, Salconi, Saleem, Salemi, Samajdar, Sanchez, Sanchez,
  Sanchis-Gual, Sanders, Santiago, Santos, Sarin, Sassolas, Sathyaprakash,
  Sauter, Savage, Savant, Sawant, Sayah, Schaetzl, Schale, Scheel, Scheuer,
  Schmidt, Schnabel, Schofield, Sch\"onbeck, Schreiber, Schulte, Schutz,
  Schwarm, Schwartz, Scott, Scott, Seidel, Sellers, Sengupta, Sennett,
  Sentenac, Sequino, Sergeev, Setyawati, Shaddock, Shaffer, Sharifi, Shahriar,
  Sharma, Sharma, Shawhan, Shen, Shikauchi, Shink, Shoemaker, Shoemaker,
  Shukla, ShyamSundar, Siellez, Sieniawska, Sigg, Singer, Singh, Singh, Singha,
  Singhal, Sintes, Sipala, Skliris, Slagmolen, Slaven-Blair, Smetana, Smith,
  Smith, Somala, Son, Soni, Sorazu, Sordini, Sorrentino, Souradeep, Sowell,
  Spencer, Spera, Srivastava, Srivastava, Staats, Stachie, Standke, Steer,
  Steinke, Steinlechner, Steinlechner, Steinmeyer, Stevenson, Stocks, Stops,
  Stover, Strain, Stratta, Strunk, Sturani, Stuver, Sudhagar, Sudhir,
  Summerscales, Sun, Sunil, Sur, Suresh, Sutton, Swinkels,
  Szczepa\ifmmode~\acute{n}\else \'{n}\fi{}czyk, Tacca, Tait, Talbot,
  Tanasijczuk, Tanner, Tao, T\'apai, Tapia, Tapia San~Martin, Tasson, Taylor,
  Tenorio, Terkowski, Thirugnanasambandam, Thomas, Thomas, Thompson, Thondapu,
  Thorne, Thrane, Tinsman, Saravanan, Tiwari, Tiwari, Tiwari, Toland, Tonelli,
  Tornasi, Torres-Forn\'e, Torrie, Tosta~e Melo, T\"oyr\"a, Travasso, Traylor,
  Tringali, Tripathee, Trovato, Trudeau, Tsang, Tse, Tso, Tsukada, Tsuna,
  Tsutsui, Turconi, Ubhi, Udall, Ueno, Ugolini, Unnikrishnan, Urban, Usman,
  Utina, Vahlbruch, Vajente, Valdes, Valentini, van Bakel, van Beuzekom,
  van~den Brand, Van Den~Broeck, Vander-Hyde, van~der Schaaf, Van~Heijningen,
  van Veggel, Vardaro, Varma, Vass, Vas\'uth, Vecchio, Vedovato, Veitch,
  Veitch, Venkateswara, Venugopalan, Verkindt, Veske, Vetrano, Vicer\'e, Viets,
  Vinciguerra, Vine, Vinet, Vitale, Vivanco, Vo, Vocca, Vorvick, Vyatchanin,
  Wade, Wade, Wade, Walet, Walker, Wallace, Wallace, Walsh, Wang, Wang, Wang,
  Ward, Warden, Warner, Was, Watchi, Weaver, Wei, Weinert, Weinstein, Weiss,
  Wellmann, Wen, We\ss{}els, Westhouse, Wette, Whelan, Whiting, Whittle,
  Wilken, Williams, Williamson, Willis, Willke, Winkler, Wipf, Wittel, Woan,
  Woehler, Wofford, Wong, Wright, Wu, Wysocki, Xiao, Yamamoto, Yang, Yang,
  Yang, Yap, Yazback, Yeeles, Yu, Yu, Yuen, Zadro\ifmmode~\dot{z}\else
  \.{z}\fi{}ny, Zadro\ifmmode~\dot{z}\else \.{z}\fi{}ny, Zanolin, Zelenova,
  Zendri, Zevin, Zhang, Zhang, Zhang, Zhao, Zhao, Zhou, Zhou, Zhu, Zimmerman,
  Zucker, \& Zweizig}]{LIGO_IMBH}
Abbott, R., Abbott, T.~D., Abraham, S., {et~al.} 2020, Phys. Rev. Lett., 125,
  101102, \dodoi{10.1103/PhysRevLett.125.101102}

\bibitem[{Alcock {et~al.}(1996)}]{MACHO_Original_results_Alcock:1995zx}
Alcock, C., {et~al.} 1996, Astrophys. J., 461, 84, \dodoi{10.1086/177039}

\bibitem[{Alcock {et~al.}(2000)Alcock, Allsman, Alves, Axelrod, Becker,
  Bennett, Cook, Drake, Freeman, Geha, Griest, Lehner, Marshall, Minniti,
  Nelson, Peterson, Popowski, Pratt, Quinn, Stubbs, Sutherland, Tomaney,
  Vandehei, \& Welch}]{Alcock_2000}
Alcock, C., Allsman, R.~A., Alves, D.~R., {et~al.} 2000, The Astrophysical
  Journal, 541, 734, \dodoi{10.1086/309484}

\bibitem[{Barranco \& Bernal(2011)}]{Barranco:2010ib}
Barranco, J., \& Bernal, A. 2011, Phys. Rev. D, 83, 043525,
  \dodoi{10.1103/PhysRevD.83.043525}

\bibitem[{Berezhiani(2004)}]{Berezhiani:2003xm}
Berezhiani, Z. 2004, Int. J. Mod. Phys., A19, 3775,
  \dodoi{10.1142/S0217751X04020075}

\bibitem[{Berezhiani {et~al.}(2006)Berezhiani, Cassisi, Ciarcelluti, \&
  Pietrinferni}]{Berezhiani:2005vv}
Berezhiani, Z., Cassisi, S., Ciarcelluti, P., \& Pietrinferni, A. 2006,
  Astropart. Phys., 24, 495, \dodoi{10.1016/j.astropartphys.2005.10.002}

\bibitem[{{Bovy}(2015)}]{starvel_Bovy_2015}
{Bovy}, J. 2015, \apjs, 216, 29, \dodoi{10.1088/0067-0049/216/2/29}

\bibitem[{Bovy(2017)}]{stardense_Bovy_2017}
Bovy, J. 2017, Monthly Notices of the Royal Astronomical Society, 470,
  1360–1387, \dodoi{10.1093/mnras/stx1277}

\bibitem[{Bovy \& Tremaine(2012)}]{localDM_Bovy:2012tw}
Bovy, J., \& Tremaine, S. 2012, Astrophys. J., 756, 89,
  \dodoi{10.1088/0004-637X/756/1/89}

\bibitem[{Bovy {et~al.}(2012)Bovy, Allende~Prieto, Beers, Bizyaev, da~Costa,
  Cunha, Ebelke, Eisenstein, Frinchaboy, García~Pérez, \&
  et~al.}]{vc_Bovy_2012}
Bovy, J., Allende~Prieto, C., Beers, T.~C., {et~al.} 2012, The Astrophysical
  Journal, 759, 131, \dodoi{10.1088/0004-637x/759/2/131}

\bibitem[{Buch {et~al.}(2019)Buch, Leung, \& Fan}]{Buch:2018qdr}
Buch, J., Leung, S. C.~J., \& Fan, J. 2019, JCAP, 04, 026,
  \dodoi{10.1088/1475-7516/2019/04/026}

\bibitem[{Burdman {et~al.}(2015)Burdman, Chacko, Harnik, de~Lima, \&
  Verhaaren}]{Burdman:2014zta}
Burdman, G., Chacko, Z., Harnik, R., de~Lima, L., \& Verhaaren, C.~B. 2015,
  Phys. Rev., D91, 055007, \dodoi{10.1103/PhysRevD.91.055007}

\bibitem[{Calcino {et~al.}(2018)Calcino, Garcia-Bellido, \&
  Davis}]{MACHO_Update_mass_formalism_Calcino:2018mwh}
Calcino, J., Garcia-Bellido, J., \& Davis, T.~M. 2018, Mon. Not. Roy. Astron.
  Soc., 479, 2889, \dodoi{10.1093/mnras/sty1368}

\bibitem[{Chacko {et~al.}(2017)Chacko, Craig, Fox, \& Harnik}]{Chacko:2016hvu}
Chacko, Z., Craig, N., Fox, P.~J., \& Harnik, R. 2017, JHEP, 07, 023,
  \dodoi{10.1007/JHEP07(2017)023}

\bibitem[{Chacko {et~al.}(2018{\natexlab{a}})Chacko, Curtin, Geller, \&
  Tsai}]{Chacko:2018vss}
Chacko, Z., Curtin, D., Geller, M., \& Tsai, Y. 2018{\natexlab{a}}, JHEP, 09,
  163, \dodoi{10.1007/JHEP09(2018)163}

\bibitem[{Chacko {et~al.}(2018{\natexlab{b}})Chacko, Curtin, Geller, \&
  Tsai}]{MTH_Cosmology_Chacko:2018vss}
---. 2018{\natexlab{b}}, JHEP, 09, 163, \dodoi{10.1007/JHEP09(2018)163}

\bibitem[{Chacko {et~al.}(2006)Chacko, Goh, \& Harnik}]{MTH_Chacko:2005pe}
Chacko, Z., Goh, H.-S., \& Harnik, R. 2006, Phys. Rev. Lett., 96, 231802,
  \dodoi{10.1103/PhysRevLett.96.231802}

\bibitem[{Chang {et~al.}(2019)Chang, Egana-Ugrinovic, Essig, \&
  Kouvaris}]{Chang:2018bgx}
Chang, J.~H., Egana-Ugrinovic, D., Essig, R., \& Kouvaris, C. 2019, JCAP, 03,
  036, \dodoi{10.1088/1475-7516/2019/03/036}

\bibitem[{Clesse \& García-Bellido(2017)}]{Clesse_2017}
Clesse, S., \& García-Bellido, J. 2017, Physics of the Dark Universe, 15,
  142–147, \dodoi{10.1016/j.dark.2016.10.002}

\bibitem[{Craig {et~al.}(2017)Craig, Koren, \& Trott}]{Craig:2016lyx}
Craig, N., Koren, S., \& Trott, T. 2017, JHEP, 05, 038,
  \dodoi{10.1007/JHEP05(2017)038}

\bibitem[{Croon {et~al.}(2020{\natexlab{a}})Croon, McKeen, \&
  Raj}]{Croon:2020wpr}
Croon, D., McKeen, D., \& Raj, N. 2020{\natexlab{a}}, Phys. Rev. D, 101,
  083013, \dodoi{10.1103/PhysRevD.101.083013}

\bibitem[{Croon {et~al.}(2020{\natexlab{b}})Croon, McKeen, Raj, \&
  Wang}]{Croon:2020ouk}
Croon, D., McKeen, D., Raj, N., \& Wang, Z. 2020{\natexlab{b}}, Phys. Rev. D,
  102, 083021, \dodoi{10.1103/PhysRevD.102.083021}

\bibitem[{Cs\'aki {et~al.}(2018)Cs\'aki, Lombardo, \& Telem}]{Csaki:2018muy}
Cs\'aki, C., Lombardo, S., \& Telem, O. 2018, {TASI Lectures on
  Non-supersymmetric BSM Models} (WSP), 501--570,
  \dodoi{10.1142/9789813233348_0007}

\bibitem[{Curtin \& Setford(2020{\natexlab{a}})}]{Curtin:2019ngc}
Curtin, D., \& Setford, J. 2020{\natexlab{a}}, JHEP, 03, 041,
  \dodoi{10.1007/JHEP03(2020)041}

\bibitem[{Curtin \& Setford(2020{\natexlab{b}})}]{Curtin:2019lhm}
---. 2020{\natexlab{b}}, Phys.\ Lett.\ B, 804, 135391,
  \dodoi{10.1016/j.physletb.2020.135391}

\bibitem[{Cyr-Racine {et~al.}(2014)Cyr-Racine, de~Putter, Raccanelli, \&
  Sigurdson}]{Cyr-Racine:2013fsa}
Cyr-Racine, F.-Y., de~Putter, R., Raccanelli, A., \& Sigurdson, K. 2014, Phys.
  Rev. D, 89, 063517, \dodoi{10.1103/PhysRevD.89.063517}

\bibitem[{Dolgov(2019)}]{dolgov2019massive}
Dolgov, A.~D. 2019, Massive Primordial Black Holes.
\newblock \doarXiv{1911.02382}

\bibitem[{{Drlica-Wagner} {et~al.}(2019){Drlica-Wagner}, {Mao}, {Adhikari},
  {Armstrong}, {Banerjee}, {Banik}, {Bechtol}, {Bird}, {Boddy}, {Bonaca},
  {Bovy}, {Buckley}, {Bulbul}, {Chang}, {Chapline}, {Cohen-Tanugi}, {Cuoco},
  {Cyr-Racine}, {Dawson}, {D{\'\i}az Rivero}, {Dvorkin}, {Erkal}, {Fassnacht},
  {Garc{\'\i}a-Bellido}, {Giannotti}, {Gluscevic}, {Golovich}, {Hendel},
  {Hezaveh}, {Horiuchi}, {Jee}, {Kaplinghat}, {Keeton}, {Koposov}, {Lam}, {Li},
  {Lu}, {Mandelbaum}, {McDermott}, {McNanna}, {Medford}, {Meyer}, {Marc},
  {Murgia}, {Nadler}, {Necib}, {Nuss}, {Pace}, {Peter}, {Polin},
  {Prescod-Weinstein}, {Read}, {Rosenfeld}, {Shipp}, {Simon}, {Slatyer},
  {Straniero}, {Strigari}, {Tollerud}, {Tyson}, {Wang}, {Wechsler}, {Wittman},
  {Yu}, {Zaharijas}, {Ali-Ha{\"\i}moud}, {Annis}, {Birrer}, {Biswas}, {Blazek},
  {Brooks}, {Buckley-Geer}, {Caputo}, {Charles}, {Digel}, {Dodelson},
  {Flaugher}, {Frieman}, {Gawiser}, {Hearin}, {Hlo{\v{z}}ek}, {Jain},
  {Jeltema}, {Koushiappas}, {Lisanti}, {LoVerde}, {Mishra-Sharma}, {Newman},
  {Nord}, {Nourbakhsh}, {Ritz}, {Robertson}, {S{\'a}nchez-Conde}, {Slosar},
  {Tait}, {Verma}, {Vilalta}, {Walter}, {Yanny}, \& {Zentner}}]{LSSTDM}
{Drlica-Wagner}, A., {Mao}, Y.-Y., {Adhikari}, S., {et~al.} 2019, arXiv
  e-prints, arXiv:1902.01055.
\newblock \doarXiv{1902.01055}

\bibitem[{Eby {et~al.}(2015)Eby, Suranyi, Vaz, \& Wijewardhana}]{Eby:2014fya}
Eby, J., Suranyi, P., Vaz, C., \& Wijewardhana, L. 2015, JHEP, 03, 080,
  \dodoi{10.1007/JHEP11(2016)134}

\bibitem[{Fan {et~al.}(2013{\natexlab{a}})Fan, Katz, Randall, \&
  Reece}]{Fan:2013yva}
Fan, J., Katz, A., Randall, L., \& Reece, M. 2013{\natexlab{a}}, Phys.Dark
  Univ., 2, 139, \dodoi{10.1016/j.dark.2013.07.001}

\bibitem[{Fan {et~al.}(2013{\natexlab{b}})Fan, Katz, Randall, \&
  Reece}]{Fan:2013tia}
---. 2013{\natexlab{b}}, Phys.Rev.Lett., 110, 211302,
  \dodoi{10.1103/PhysRevLett.110.211302}

\bibitem[{Foot(1999)}]{DM_historical1}
Foot, R. 1999, Phys. Lett., B452, 83, \dodoi{10.1016/S0370-2693(99)00230-0}

\bibitem[{Foot {et~al.}(2001)Foot, Ignatiev, \& Volkas}]{Foot:2000vy}
Foot, R., Ignatiev, A.~{\relax Yu}., \& Volkas, R.~R. 2001, Phys. Lett., B503,
  355, \dodoi{10.1016/S0370-2693(01)00228-3}

\bibitem[{Golovich {et~al.}(2020)Golovich, Dawson, Bartolić, Lam, Lu, Medford,
  Schneider, Chapline, Schlafly, Drlica-Wagner, \&
  Pruett}]{golovich2020reanalysis}
Golovich, N., Dawson, W.~A., Bartolić, F., {et~al.} 2020, A Reanalysis of
  Public Galactic Bulge Gravitational Microlensing Events from OGLE-III and IV.
\newblock \doarXiv{2009.07927}

\bibitem[{{Gould}(1992)}]{Gould92a}
{Gould}, A. 1992, \apj, 392, 442, \dodoi{10.1086/171443}

\bibitem[{{Griest}(1991)}]{griest_originalderivation}
{Griest}, K. 1991, \apj, 366, 412, \dodoi{10.1086/169575}

\bibitem[{Hogan \& Rees(1988)}]{Hogan:1988mp}
Hogan, C., \& Rees, M. 1988, Phys.Lett.B, 205, 228,
  \dodoi{10.1016/0370-2693(88)91655-3}

\bibitem[{{Ivezi{\'c}} {et~al.}(2019){Ivezi{\'c}}, {Kahn}, {Tyson}, {Abel},
  {Acosta}, {Allsman}, {Alonso}, {AlSayyad}, {Anderson}, {Andrew}, {Angel},
  {Angeli}, {Ansari}, {Antilogus}, {Araujo}, {Armstrong}, {Arndt}, {Astier},
  {Aubourg}, {Auza}, {Axelrod}, {Bard}, {Barr}, {Barrau}, {Bartlett}, {Bauer},
  {Bauman}, {Baumont}, {Bechtol}, {Bechtol}, {Becker}, {Becla}, {Beldica},
  {Bellavia}, {Bianco}, {Biswas}, {Blanc}, {Blazek}, {Bland ford}, {Bloom},
  {Bogart}, {Bond}, {Booth}, {Borgland}, {Borne}, {Bosch}, {Boutigny},
  {Brackett}, {Bradshaw}, {Brand t}, {Brown}, {Bullock}, {Burchat}, {Burke},
  {Cagnoli}, {Calabrese}, {Callahan}, {Callen}, {Carlin}, {Carlson}, {Chand
  rasekharan}, {Charles-Emerson}, {Chesley}, {Cheu}, {Chiang}, {Chiang},
  {Chirino}, {Chow}, {Ciardi}, {Claver}, {Cohen-Tanugi}, {Cockrum}, {Coles},
  {Connolly}, {Cook}, {Cooray}, {Covey}, {Cribbs}, {Cui}, {Cutri}, {Daly},
  {Daniel}, {Daruich}, {Daubard}, {Daues}, {Dawson}, {Delgado}, {Dellapenna},
  {de Peyster}, {de Val-Borro}, {Digel}, {Doherty}, {Dubois},
  {Dubois-Felsmann}, {Durech}, {Economou}, {Eifler}, {Eracleous}, {Emmons},
  {Fausti Neto}, {Ferguson}, {Figueroa}, {Fisher-Levine}, {Focke}, {Foss},
  {Frank}, {Freemon}, {Gangler}, {Gawiser}, {Geary}, {Gee}, {Geha}, {Gessner},
  {Gibson}, {Gilmore}, {Glanzman}, {Glick}, {Goldina}, {Goldstein}, {Goodenow},
  {Graham}, {Gressler}, {Gris}, {Guy}, {Guyonnet}, {Haller}, {Harris},
  {Hascall}, {Haupt}, {Hernand ez}, {Herrmann}, {Hileman}, {Hoblitt},
  {Hodgson}, {Hogan}, {Howard}, {Huang}, {Huffer}, {Ingraham}, {Innes},
  {Jacoby}, {Jain}, {Jammes}, {Jee}, {Jenness}, {Jernigan}, {Jevremovi{\'c}},
  {Johns}, {Johnson}, {Johnson}, {Jones}, {Juramy-Gilles}, {Juri{\'c}},
  {Kalirai}, {Kallivayalil}, {Kalmbach}, {Kantor}, {Karst}, {Kasliwal},
  {Kelly}, {Kessler}, {Kinnison}, {Kirkby}, {Knox}, {Kotov}, {Krabbendam},
  {Krughoff}, {Kub{\'a}nek}, {Kuczewski}, {Kulkarni}, {Ku}, {Kurita}, {Lage},
  {Lambert}, {Lange}, {Langton}, {Le Guillou}, {Levine}, {Liang}, {Lim},
  {Lintott}, {Long}, {Lopez}, {Lotz}, {Lupton}, {Lust}, {MacArthur}, {Mahabal},
  {Mand elbaum}, {Markiewicz}, {Marsh}, {Marshall}, {Marshall}, {May},
  {McKercher}, {McQueen}, {Meyers}, {Migliore}, {Miller}, {Mills}, {Miraval},
  {Moeyens}, {Moolekamp}, {Monet}, {Moniez}, {Monkewitz}, {Montgomery},
  {Morrison}, {Mueller}, {Muller}, {Mu{\~n}oz Arancibia}, {Neill}, {Newbry},
  {Nief}, {Nomerotski}, {Nordby}, {O'Connor}, {Oliver}, {Olivier}, {Olsen},
  {O'Mullane}, {Ortiz}, {Osier}, {Owen}, {Pain}, {Palecek}, {Parejko},
  {Parsons}, {Pease}, {Peterson}, {Peterson}, {Petravick}, {Libby Petrick},
  {Petry}, {Pierfederici}, {Pietrowicz}, {Pike}, {Pinto}, {Plante}, {Plate},
  {Plutchak}, {Price}, {Prouza}, {Radeka}, {Rajagopal}, {Rasmussen},
  {Regnault}, {Reil}, {Reiss}, {Reuter}, {Ridgway}, {Riot}, {Ritz}, {Robinson},
  {Roby}, {Roodman}, {Rosing}, {Roucelle}, {Rumore}, {Russo}, {Saha},
  {Sassolas}, {Schalk}, {Schellart}, {Schindler}, {Schmidt}, {Schneider},
  {Schneider}, {Schoening}, {Schumacher}, {Schwamb}, {Sebag}, {Selvy},
  {Sembroski}, {Seppala}, {Serio}, {Serrano}, {Shaw}, {Shipsey}, {Sick},
  {Silvestri}, {Slater}, {Smith}, {Smith}, {Sobhani}, {Soldahl},
  {Storrie-Lombardi}, {Stover}, {Strauss}, {Street}, {Stubbs}, {Sullivan},
  {Sweeney}, {Swinbank}, {Szalay}, {Takacs}, {Tether}, {Thaler}, {Thayer},
  {Thomas}, {Thornton}, {Thukral}, {Tice}, {Trilling}, {Turri}, {Van Berg},
  {Vanden Berk}, {Vetter}, {Virieux}, {Vucina}, {Wahl}, {Walkowicz}, {Walsh},
  {Walter}, {Wang}, {Wang}, {Warner}, {Wiecha}, {Willman}, {Winters},
  {Wittman}, {Wolff}, {Wood-Vasey}, {Wu}, {Xin}, {Yoachim}, \&
  {Zhan}}]{LSST_specs}
{Ivezi{\'c}}, {\v{Z}}., {Kahn}, S.~M., {Tyson}, J.~A., {et~al.} 2019, \apj,
  873, 111, \dodoi{10.3847/1538-4357/ab042c}

\bibitem[{Ivezić(2016)}]{ivezic_2016}
Ivezić, Z. 2016, Proceedings of the International Astronomical Union, 12,
  330–337, \dodoi{10.1017/S1743921316012424}

\bibitem[{Ivezić {et~al.}(2019)Ivezić, Kahn, Tyson, Abel, Acosta, Allsman,
  Alonso, AlSayyad, Anderson, Andrew, \& et~al.}]{Ivezi__2019}
Ivezić, Z., Kahn, S.~M., Tyson, J.~A., {et~al.} 2019, The Astrophysical
  Journal, 873, 111, \dodoi{10.3847/1538-4357/ab042c}

\bibitem[{Jedamzik(2020)}]{Jedamzik:2020ypm}
Jedamzik, K. 2020, JCAP, 09, 022, \dodoi{10.1088/1475-7516/2020/09/022}

\bibitem[{Johnson {et~al.}(2020)Johnson, Penny, Gaudi, Kerins, Rattenbury,
  Robin, Calchi~Novati, \& Henderson}]{Johnson_2020}
Johnson, S.~A., Penny, M., Gaudi, B.~S., {et~al.} 2020, The Astronomical
  Journal, 160, 123, \dodoi{10.3847/1538-3881/aba75b}

\bibitem[{Kolb \& Tkachev(1993)}]{Kolb:1993zz}
Kolb, E.~W., \& Tkachev, I.~I. 1993, Phys. Rev. Lett., 71, 3051,
  \dodoi{10.1103/PhysRevLett.71.3051}

\bibitem[{Kramer \& Randall(2016{\natexlab{a}})}]{Kramer:2016dew}
Kramer, E.~D., \& Randall, L. 2016{\natexlab{a}}, Astrophys.J., 829, 126,
  \dodoi{10.3847/0004-637X/829/2/126}

\bibitem[{Kramer \& Randall(2016{\natexlab{b}})}]{Kramer:2016dqu}
---. 2016{\natexlab{b}}, Astrophys.J., 824, 116,
  \dodoi{10.3847/0004-637X/824/2/116}

\bibitem[{{Kroupa}(2001)}]{Kroupa_IMF}
{Kroupa}, P. 2001, \mnras, 322, 231, \dodoi{10.1046/j.1365-8711.2001.04022.x}

\bibitem[{{LSST Science Collaboration} {et~al.}(2009){LSST Science
  Collaboration}, {Abell}, {Allison}, {Anderson}, {Andrew}, {Angel}, {Armus},
  {Arnett}, {Asztalos}, {Axelrod}, {Bailey}, {Ballantyne}, {Bankert},
  {Barkhouse}, {Barr}, {Barrientos}, {Barth}, {Bartlett}, {Becker}, {Becla},
  {Beers}, {Bernstein}, {Biswas}, {Blanton}, {Bloom}, {Bochanski}, {Boeshaar},
  {Borne}, {Bradac}, {Brandt}, {Bridge}, {Brown}, {Brunner}, {Bullock},
  {Burgasser}, {Burge}, {Burke}, {Cargile}, {Chandrasekharan}, {Chartas},
  {Chesley}, {Chu}, {Cinabro}, {Claire}, {Claver}, {Clowe}, {Connolly}, {Cook},
  {Cooke}, {Cooray}, {Covey}, {Culliton}, {de Jong}, {de Vries}, {Debattista},
  {Delgado}, {Dell'Antonio}, {Dhital}, {Di Stefano}, {Dickinson}, {Dilday},
  {Djorgovski}, {Dobler}, {Donalek}, {Dubois-Felsmann}, {Durech},
  {Eliasdottir}, {Eracleous}, {Eyer}, {Falco}, {Fan}, {Fassnacht}, {Ferguson},
  {Fernandez}, {Fields}, {Finkbeiner}, {Figueroa}, {Fox}, {Francke}, {Frank},
  {Frieman}, {Fromenteau}, {Furqan}, {Galaz}, {Gal-Yam}, {Garnavich},
  {Gawiser}, {Geary}, {Gee}, {Gibson}, {Gilmore}, {Grace}, {Green}, {Gressler},
  {Grillmair}, {Habib}, {Haggerty}, {Hamuy}, {Harris}, {Hawley}, {Heavens},
  {Hebb}, {Henry}, {Hileman}, {Hilton}, {Hoadley}, {Holberg}, {Holman},
  {Howell}, {Infante}, {Ivezic}, {Jacoby}, {Jain}, {R}, {Jedicke}, {Jee},
  {Garrett Jernigan}, {Jha}, {Johnston}, {Jones}, {Juric}, {Kaasalainen},
  {Styliani}, {Kafka}, {Kahn}, {Kaib}, {Kalirai}, {Kantor}, {Kasliwal},
  {Keeton}, {Kessler}, {Knezevic}, {Kowalski}, {Krabbendam}, {Krughoff},
  {Kulkarni}, {Kuhlman}, {Lacy}, {Lepine}, {Liang}, {Lien}, {Lira}, {Long},
  {Lorenz}, {Lotz}, {Lupton}, {Lutz}, {Macri}, {Mahabal}, {Mandelbaum},
  {Marshall}, {May}, {McGehee}, {Meadows}, {Meert}, {Milani}, {Miller},
  {Miller}, {Mills}, {Minniti}, {Monet}, {Mukadam}, {Nakar}, {Neill}, {Newman},
  {Nikolaev}, {Nordby}, {O'Connor}, {Oguri}, {Oliver}, {Olivier}, {Olsen},
  {Olsen}, {Olszewski}, {Oluseyi}, {Padilla}, {Parker}, {Pepper}, {Peterson},
  {Petry}, {Pinto}, {Pizagno}, {Popescu}, {Prsa}, {Radcka}, {Raddick},
  {Rasmussen}, {Rau}, {Rho}, {Rhoads}, {Richards}, {Ridgway}, {Robertson},
  {Roskar}, {Saha}, {Sarajedini}, {Scannapieco}, {Schalk}, {Schindler},
  {Schmidt}, {Schmidt}, {Schneider}, {Schumacher}, {Scranton}, {Sebag},
  {Seppala}, {Shemmer}, {Simon}, {Sivertz}, {Smith}, {Allyn Smith}, {Smith},
  {Spitz}, {Stanford}, {Stassun}, {Strader}, {Strauss}, {Stubbs}, {Sweeney},
  {Szalay}, {Szkody}, {Takada}, {Thorman}, {Trilling}, {Trimble}, {Tyson}, {Van
  Berg}, {Vanden Berk}, {VanderPlas}, {Verde}, {Vrsnak}, {Walkowicz},
  {Wandelt}, {Wang}, {Wang}, {Warner}, {Wechsler}, {West}, {Wiecha},
  {Williams}, {Willman}, {Wittman}, {Wolff}, {Wood-Vasey}, {Wozniak}, {Young},
  {Zentner}, \& {Zhan}}]{LSSTScieneBook}
{LSST Science Collaboration}, {Abell}, P.~A., {Allison}, J., {et~al.} 2009,
  arXiv e-prints, arXiv:0912.0201.
\newblock \doarXiv{0912.0201}

\bibitem[{Lu {et~al.}(2019)Lu, Lam, Medford, Dawson, \&
  Golovich}]{PBH_EinsteinTime_Lu:2019hoc}
Lu, J.~R., Lam, C.~Y., Medford, M., Dawson, W., \& Golovich, N. 2019, arXiv
  e-prints, \dodoi{10.3847/2515-5172/ab1421}

\bibitem[{Martin(2010)}]{Martin:1997ns}
Martin, S.~P. 2010, Adv. Ser. Direct. High Energy Phys., 21, 1,
  \dodoi{10.1142/9789812839657_0001}

\bibitem[{Mohapatra \& Teplitz(1997)}]{Mohapatra:1996yy}
Mohapatra, R.~N., \& Teplitz, V.~L. 1997, Astrophys. J., 478, 29,
  \dodoi{10.1086/303762}

\bibitem[{Mohapatra \& Teplitz(1999)}]{Mohapatra:1999ih}
---. 1999, Phys. Lett., B462, 302, \dodoi{10.1016/S0370-2693(99)00789-3}

\bibitem[{Navarro {et~al.}(1996)Navarro, Frenk, \& White}]{Navarro_1996}
Navarro, J.~F., Frenk, C.~S., \& White, S. D.~M. 1996, The Astrophysical
  Journal, 462, 563, \dodoi{10.1086/177173}

\bibitem[{Nesti \& Salucci(2013)}]{MW_halo_dist}
Nesti, F., \& Salucci, P. 2013, JCAP, 1307, 016,
  \dodoi{10.1088/1475-7516/2013/07/016}

\bibitem[{Niikura {et~al.}(2019)}]{Niikura:2017zjd}
Niikura, H., {et~al.} 2019, Nature Astron., 3, 524,
  \dodoi{10.1038/s41550-019-0723-1}

\bibitem[{Paczynski(1986)}]{ML_Original_Paczynski:1985jf}
Paczynski, B. 1986, Astrophys. J., 304, 1, \dodoi{10.1086/164140}

\bibitem[{{Rich}(2018)}]{lsst_resolution}
{Rich}, R.~M. 2018, in IAU Symposium, Vol. 334, Rediscovering Our Galaxy, ed.
  C.~{Chiappini}, I.~{Minchev}, E.~{Starkenburg}, \& M.~{Valentini}, 233--241,
  \dodoi{10.1017/S1743921317009413}

\bibitem[{Rich {et~al.}(2020)Rich, Johnson, Young, Simion, Clarkson,
  Pilachowski, Michael, Kunder, Vivas, Koch, Marchetti, Ibata, Martin, Robin,
  Lagarde, Collins, Ivezi{\'{c}}, de~Propris, Shen, Gerhard, \&
  Soto}]{Rich_2020}
Rich, R.~M., Johnson, C.~I., Young, M., {et~al.} 2020, Monthly Notices of the
  Royal Astronomical Society, 499, 2340, \dodoi{10.1093/mnras/staa2426}

\bibitem[{{Sajadian} \& {Poleski}(2019)}]{LSST_MLPredictions}
{Sajadian}, S., \& {Poleski}, R. 2019, \apj, 871, 205,
  \dodoi{10.3847/1538-4357/aafa1d}

\bibitem[{Sajadian \& Poleski(2019)}]{Sajadian_2019}
Sajadian, S., \& Poleski, R. 2019, The Astrophysical Journal, 871, 205,
  \dodoi{10.3847/1538-4357/aafa1d}

\bibitem[{Schlafly {et~al.}(2018)Schlafly, Green, Lang, Daylan, Finkbeiner,
  Lee, Meisner, Schlegel, \& Valdes}]{Schlafly_2018}
Schlafly, E.~F., Green, G.~M., Lang, D., {et~al.} 2018, The Astrophysical
  Journal Supplement Series, 234, 39, \dodoi{10.3847/1538-4365/aaa3e2}

\bibitem[{Schutz {et~al.}(2018{\natexlab{a}})Schutz, Lin, Safdi, \&
  Wu}]{Gaia_thindisk}
Schutz, K., Lin, T., Safdi, B.~R., \& Wu, C.-L. 2018{\natexlab{a}}, Physical
  Review Letters, 121, \dodoi{10.1103/physrevlett.121.081101}

\bibitem[{Schutz {et~al.}(2018{\natexlab{b}})Schutz, Lin, Safdi, \&
  Wu}]{Schutz:2017tfp}
---. 2018{\natexlab{b}}, Phys. Rev. Lett., 121, 081101,
  \dodoi{10.1103/PhysRevLett.121.081101}

\bibitem[{{Sharma} {et~al.}(2011){Sharma}, {Bland-Hawthorn}, {Johnston}, \&
  {Binney}}]{GALAXIA_reference}
{Sharma}, S., {Bland-Hawthorn}, J., {Johnston}, K.~V., \& {Binney}, J. 2011,
  \apj, 730, 3, \dodoi{10.1088/0004-637X/730/1/3}

\bibitem[{{Skowron} {et~al.}(2015){Skowron}, {Shin}, {Udalski}, {Han}, {Sumi},
  {Shvartzvald}, {Gould}, {Dominis Prester}, {Street}, {J{\o}rgensen},
  {Bennett}, {Bozza}, {Szyma{\'n}ski}, {Kubiak}, {Pietrzy{\'n}ski},
  {Soszy{\'n}ski}, {Poleski}, {Koz{\l}owski}, {Pietrukowicz}, {Ulaczyk},
  {Wyrzykowski}, {OGLE Collaboration}, {Abe}, {Bhattacharya}, {Bond},
  {Botzler}, {Freeman}, {Fukui}, {Fukunaga}, {Itow}, {Ling}, {Koshimoto},
  {Masuda}, {Matsubara}, {Muraki}, {Namba}, {Ohnishi}, {Philpott},
  {Rattenbury}, {Saito}, {Sullivan}, {Suzuki}, {Tristram}, {Yock}, {MOA
  Collaboration}, {Maoz}, {Kaspi}, {Friedmann}, {Wise Group}, {Almeida},
  {Batista}, {Christie}, {Choi}, {DePoy}, {Gaudi}, {Henderson}, {Hwang},
  {Jablonski}, {Jung}, {Lee}, {McCormick}, {Natusch}, {Ngan}, {Park}, {Pogge},
  {Yee}, {{\ensuremath{\mu}}FUN Collaboration}, {Albrow}, {Bachelet},
  {Beaulieu}, {Brillant}, {Caldwell}, {Cassan}, {Cole}, {Corrales}, {Coutures},
  {Dieters}, {Donatowicz}, {Fouqu{\'e}}, {Greenhill}, {Kains}, {Kane}, {Kubas},
  {Marquette}, {Martin}, {Menzies}, {Pollard}, {Ranc}, {Sahu}, {Wambsganss},
  {Williams}, {Wouters}, {PLANET Collaboration}, {Tsapras}, {Bramich}, {Horne},
  {Hundertmark}, {Snodgrass}, {Steele}, {RoboNet Collaboration}, {Alsubai},
  {Browne}, {Burgdorf}, {Calchi Novati}, {Dodds}, {Dominik}, {Dreizler},
  {Fang}, {Gu}, {Hardis}, {Harps{\o}e}, {Hessman}, {Hinse}, {Hornstrup},
  {Jessen-Hansen}, {Kerins}, {Liebig}, {Lund}, {Lundkvist}, {Mancini},
  {Mathiasen}, {Penny}, {Rahvar}, {Ricci}, {Scarpetta}, {Skottfelt},
  {Southworth}, {Surdej}, {Tregloan-Reed}, {Wertz}, \& {MiNDSTEp
  Consortium}}]{planet_mL_1}
{Skowron}, J., {Shin}, I.~G., {Udalski}, A., {et~al.} 2015, \apj, 804, 33,
  \dodoi{10.1088/0004-637X/804/1/33}

\bibitem[{{Skowron} {et~al.}(2018){Skowron}, {Ryu}, {Hwang}, {Udalski},
  {Mr{\'o}z}, {Koz{\l}owski}, {Soszy{\'n}ski}, {Pietrukowicz}, {Szyma{\'n}ski},
  {Poleski}, {Ulaczyk}, {Pawlak}, {Rybicki}, {Iwanek}, {Albrow}, {Chung},
  {Gould}, {Han}, {Jung}, {Shin}, {Shvartzvald}, {Yee}, {Zang}, {Zhu}, {Cha},
  {Kim}, {Kim}, {Kim}, {Lee}, {Lee}, {Lee}, {Park}, \& {Pogge}}]{planet_mL}
{Skowron}, J., {Ryu}, Y.~H., {Hwang}, K.~H., {et~al.} 2018, Acta Astronomica,
  68, 43, \dodoi{10.32023/0001-5237/68.1.2}

\bibitem[{{The GRAVITY Collaboration} {et~al.}(2019){The GRAVITY
  Collaboration}, {Abuter, R.}, {Amorim, A.}, {Baub\"ock, M.}, {Berger, J. P.},
  {Bonnet, H.}, {Brandner, W.}, {Cl\'enet, Y.}, {Coud\'e du Foresto, V.}, {de
  Zeeuw, P. T.}, {Dexter, J.}, {Duvert, G.}, {Eckart, A.}, {Eisenhauer, F.},
  {F\"orster Schreiber, N. M.}, {Garcia, P.}, {Gao, F.}, {Gendron, E.},
  {Genzel, R.}, {Gerhard, O.}, {Gillessen, S.}, {Habibi, M.}, {Haubois, X.},
  {Henning, T.}, {Hippler, S.}, {Horrobin, M.}, {Jim\'enez-Rosales, A.},
  {Jocou, L.}, {Kervella, P.}, {Lacour, S.}, {Lapeyr\`ere, V.}, {Le Bouquin,
  J.-B.}, {L\'ena, P.}, {Ott, T.}, {Paumard, T.}, {Perraut, K.}, {Perrin, G.},
  {Pfuhl, O.}, {Rabien, S.}, {Rodriguez Coira, G.}, {Rousset, G.},
  {Scheithauer, S.}, {Sternberg, A.}, {Straub, O.}, {Straubmeier, C.}, {Sturm,
  E.}, {Tacconi, L. J.}, {Vincent, F.}, {von Fellenberg, S.}, {Waisberg, I.},
  {Widmann, F.}, {Wieprecht, E.}, {Wiezorrek, E.}, {Woillez, J.}, \& {Yazici,
  S.}}]{GRAVITY}
{The GRAVITY Collaboration}, {Abuter, R.}, {Amorim, A.}, {et~al.} 2019, A\&A,
  625, L10, \dodoi{10.1051/0004-6361/201935656}

\bibitem[{{Tisserand} {et~al.}(2007){Tisserand}, {Le Guillou, L.}, {Afonso,
  C.}, {Albert, J. N.}, {Andersen, J.}, {Ansari, R.}, {Aubourg, \'E.},
  {Bareyre, P.}, {Beaulieu, J. P.}, {Charlot, X.}, {Coutures, C.}, {Ferlet,
  R.}, {Fouqu\'e, P.}, {Glicenstein, J. F.}, {Goldman, B.}, {Gould, A.},
  {Graff, D.}, {Gros, M.}, {Haissinski, J.}, {Hamadache, C.}, {de Kat, J.},
  {Lasserre, T.}, {Lesquoy, \'E.}, {Loup, C.}, {Magneville, C.}, {Marquette, J.
  B.}, {Maurice, \'E.}, {Maury, A.}, {Milsztajn, A.}, {Moniez, M.},
  {Palanque-Delabrouille, N.}, {Perdereau, O.}, {Rahal, Y. R.}, {Rich, J.},
  {Spiro, M.}, {Vidal-Madjar, A.}, {Vigroux, L.}, \& {S. Zylberajch (The EROS-2
  collaboration)}}]{Tisserand}
{Tisserand}, {Le Guillou, L.}, {Afonso, C.}, {et~al.} 2007, A\&A, 469, 387,
  \dodoi{10.1051/0004-6361:20066017}

\bibitem[{Tkachev(1991)}]{Tkachev:1991ka}
Tkachev, I. 1991, Phys.Lett.B, 261, 289, \dodoi{10.1016/0370-2693(91)90330-S}

\bibitem[{{Wegg} {et~al.}(2016){Wegg}, {Gerhard}, \&
  {Portail}}]{ML_Bulge_2016MNRAS.463..557W}
{Wegg}, C., {Gerhard}, O., \& {Portail}, M. 2016, \mnras, 463, 557,
  \dodoi{10.1093/mnras/stw1954}

\bibitem[{Winch(2020)}]{harrisongithub}
Winch, H. 2020, {https://github.com/HarrisonWinch96/DarkDisk\_Microlensing}

\bibitem[{Wyrzykowski {et~al.}(2009)Wyrzykowski, Kozlowski, Skowron, Belokurov,
  Smith, Udalski, Szymański, Kubiak, Pietrzyński, Soszyński, Szewczyk, \&
  Żebruń}]{Wyrzykowski_2009}
Wyrzykowski, L., Kozlowski, S., Skowron, J., {et~al.} 2009, Monthly Notices of
  the Royal Astronomical Society, 397, 1228,
  \dodoi{10.1111/j.1365-2966.2009.15029.x}

\bibitem[{Wyrzykowski
  {et~al.}(2016{\natexlab{a}})}]{OGLE_ML_Wyrzykowski:2015ppa}
Wyrzykowski, L., {et~al.} 2016{\natexlab{a}}, Mon. Not. Roy. Astron. Soc., 458,
  3012, \dodoi{10.1093/mnras/stw426}

\bibitem[{Wyrzykowski {et~al.}(2016{\natexlab{b}})Wyrzykowski,
  Kostrzewa-Rutkowska, Skowron, Rybicki, Mróz, Kozłowski, Udalski,
  Szymański, Pietrzyński, Soszyński, \& et~al.}]{Wyrzykowski_2016}
Wyrzykowski, L., Kostrzewa-Rutkowska, Z., Skowron, J., {et~al.}
  2016{\natexlab{b}}, Monthly Notices of the Royal Astronomical Society, 458,
  3012–3026, \dodoi{10.1093/mnras/stw426}

\end{thebibliography}

\end{document}